\documentclass[11pt]{paper}
\usepackage{graphicx}

\usepackage{geometry}                
\usepackage{graphicx}

\usepackage{pdflscape}

\usepackage[usenames,dvipsnames]{xcolor}
\usepackage{amssymb}
\newtheorem{definition}{Definition}
\usepackage{calc,url,xspace}
\usepackage{booktabs}
\usepackage{rotating}
\usepackage[hidelinks=true]{hyperref}
\usepackage{wrapfig}

\usepackage[numbers]{natbib}

\usepackage{times}

\newcommand{\commentout}[1]{}

\newcommand{\eg}{e.g.,\xspace}
\newcommand{\ie}{i.e.,\xspace}
\newcommand{\etc}{etc\xspace}

\newcommand{\commented}[1]{\textsf{\textbf{\LARGE #1}}\typeout{++ #1}}

\newcommand{\later}[1]{{\color{Red} \commented{LATER: #1}}}
\renewcommand{\later}[1]{}

\definecolor{DarkGray}{HTML}{808080}

\newcommand{\defn}[1]{\emph{#1}}
\newcommand{\sysname}[1]{\emph{#1}}

\def\toplayer{\textsc{Principles Layer\xspace}}
\def\midlayer{\textsc{Rules Layer\xspace}}
\def\lowlayer{\textsc{Reactions Layer\xspace}}

\title{Towards a Framework for Certification\\ of Reliable Autonomous Systems}

\author{
  \textbf{Michael Fisher}\\
  Department of Computer Science, University of Liverpool, United Kingdom \\
  \texttt{mfisher@liverpool.ac.uk}
  \and
  \textbf{Viviana Mascardi}\\
  Department of Informatics, Bioengineering, Robotics, \&{} Systems Engineering,\\
  University of Genova, Italy \\
  \texttt{viviana.mascardi@unige.it}
  \and
  \textbf{Kristin Yvonne Rozier}\\
  Iowa State University, Iowa, USA \\
  \texttt{kyrozier@iastate.edu}
  \and
  \textbf{Bernd-Holger Schlingloff}\\
  Humboldt University and Fraunhofer FOKUS, Berlin, Germany\\
  \texttt{hs@informatik.hu-berlin.de}
  \and
  \textbf{Michael Winikoff}\thanks{Michael Winikoff was at the University of Otago when part of this work was done.}\\
  Victoria University of Wellington, New Zealand  \\
   \texttt{winikoff@gmail.com}
   \and
   \textbf{Neil Yorke-Smith}\\
   Delft University of Technology, The Netherlands,\\ and
              American University of Beirut, Lebanon \\
              \texttt{n.yorke-smith@tudelft.nl}
              }

\date{January 2020}                                           

\begin{document}

\maketitle
\begin{center}
  {\Large \fbox{This paper is under review for journal publication.}}
  \vspace*{2em}
\end{center}
\vfill

\newpage
\begin{abstract}
A computational system is called \emph{autonomous} if it is able to make its own
decisions, or take its own actions, without human supervision or
control.  The capability and spread of such systems have reached the
point where they are beginning to touch much of everyday life.
However, regulators grapple with how to deal with autonomous systems,
for example how could we certify an Unmanned Aerial System for autonomous
use in civilian airspace?  We here analyse what is needed in order to
provide verified reliable behaviour of an autonomous system, analyse
what can be done as the state-of-the-art in automated verification,
and propose a roadmap towards developing regulatory guidelines, 
including articulating challenges to researchers, to engineers, and to regulators.  
Case studies in seven distinct domains illustrate the article.
%
\end{abstract}

\section{Introduction}
\label{sec:intro}



Since the dawn of human history, humans have designed, implemented and
adopted tools to make it easier to perform tasks, often improving
efficiency, safety, or security.
Indeed, recent studies
show a direct relationship between increasing technological
complexity, cognitive evolution and cultural variation~\citep{stout2011stone}.

\begin{figure}[htb]
\begin{center}
\includegraphics[trim={2cm 1.5cm 2.6cm 3cm},clip,height=5cm]{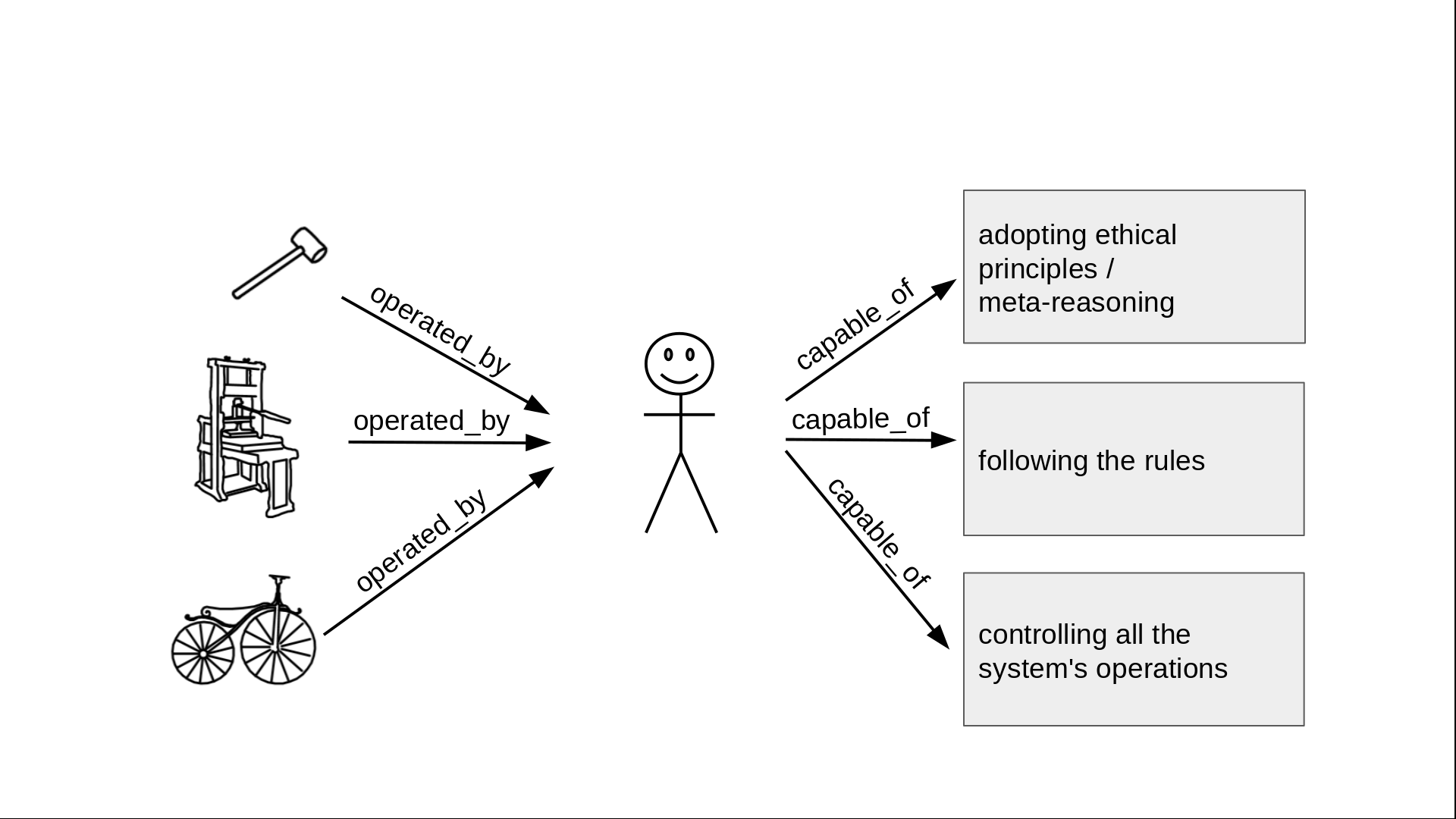}
\caption{Human user has the full control of the tool.}
\label{fig:img1}
\end{center}
\end{figure}

When such tools were simple, the person using the tool had full
control over the way the tool should be operated, understood why it
worked in that way, knew how the tool should be used to comply with
existing rules, and when such rules might be broken if the situation
demanded an exceptional use of the tool.  For example, our early
ancestors could use a hammer for building artefacts, knew why the
hammer could be used for their purposes, followed the rules of not using
it as a weapon against other humans, but might have chosen to break
this rule if their families were in danger (Figure~\ref{fig:img1}).

\begin{figure}[h]
\begin{center}
\includegraphics[trim={2cm 1.5cm 2.6cm 3cm},clip,height=5cm]{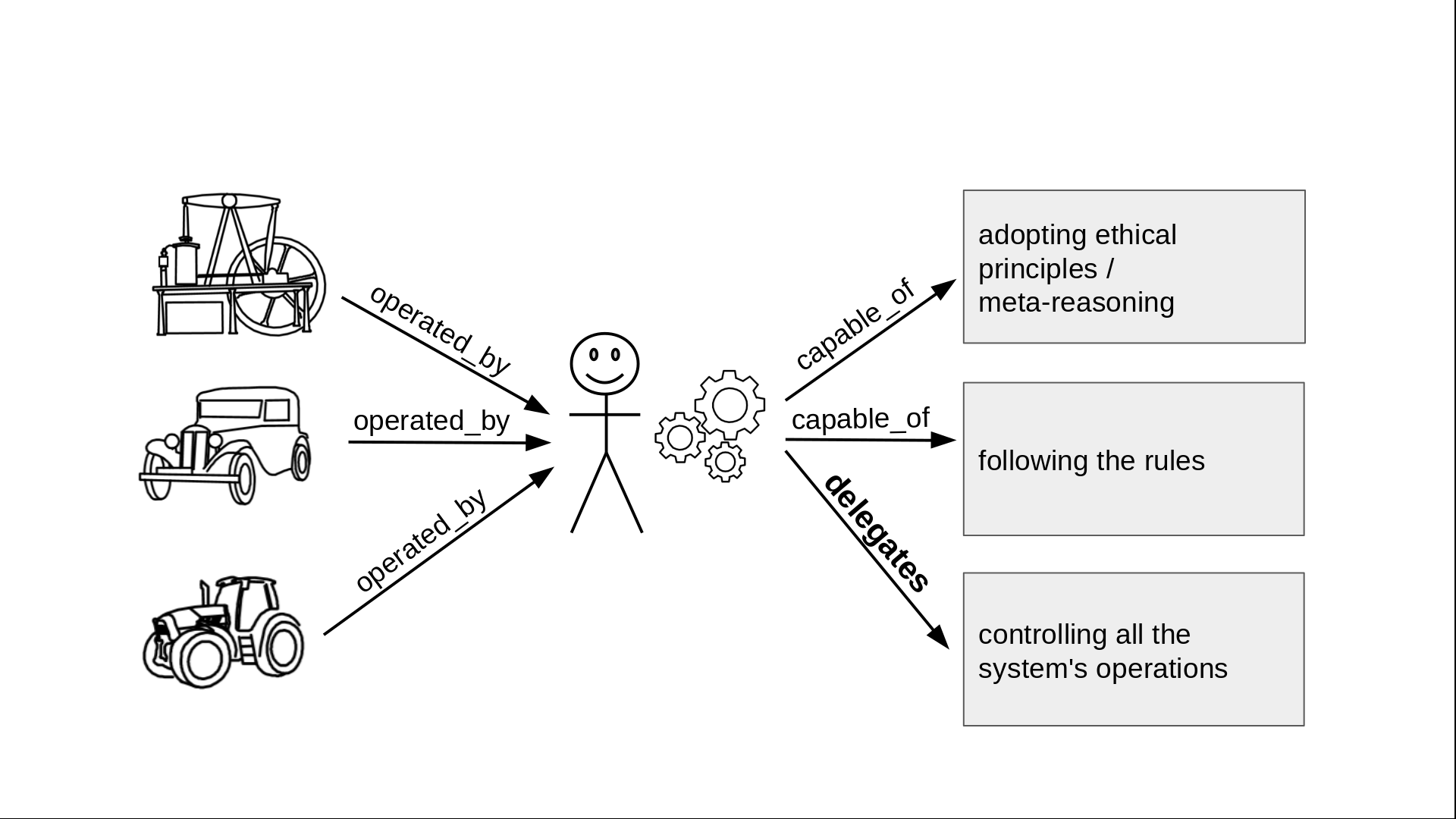}
\caption{Human (partially) delegates the control of the operations.}
\label{fig:img2}
\end{center}
\end{figure}

However, as tools became more complex and developed into systems 
composed of many different parts, 
users lost their broad view on
how the system, or even some of its components, worked and -- without that
know-how -- they lost part of their control over the system.  But
users still retained the capability of using systems following the
rules, and breaking the rules if needed.  By delegating the control of
some basic tasks
to the system itself, users gained in efficiency at the expense of
exhaustive control (Figure~\ref{fig:img2}).

\begin{figure}[h!]
\begin{center}
\includegraphics[trim={2cm 1.5cm 2.6cm 3cm},clip,height=5cm]{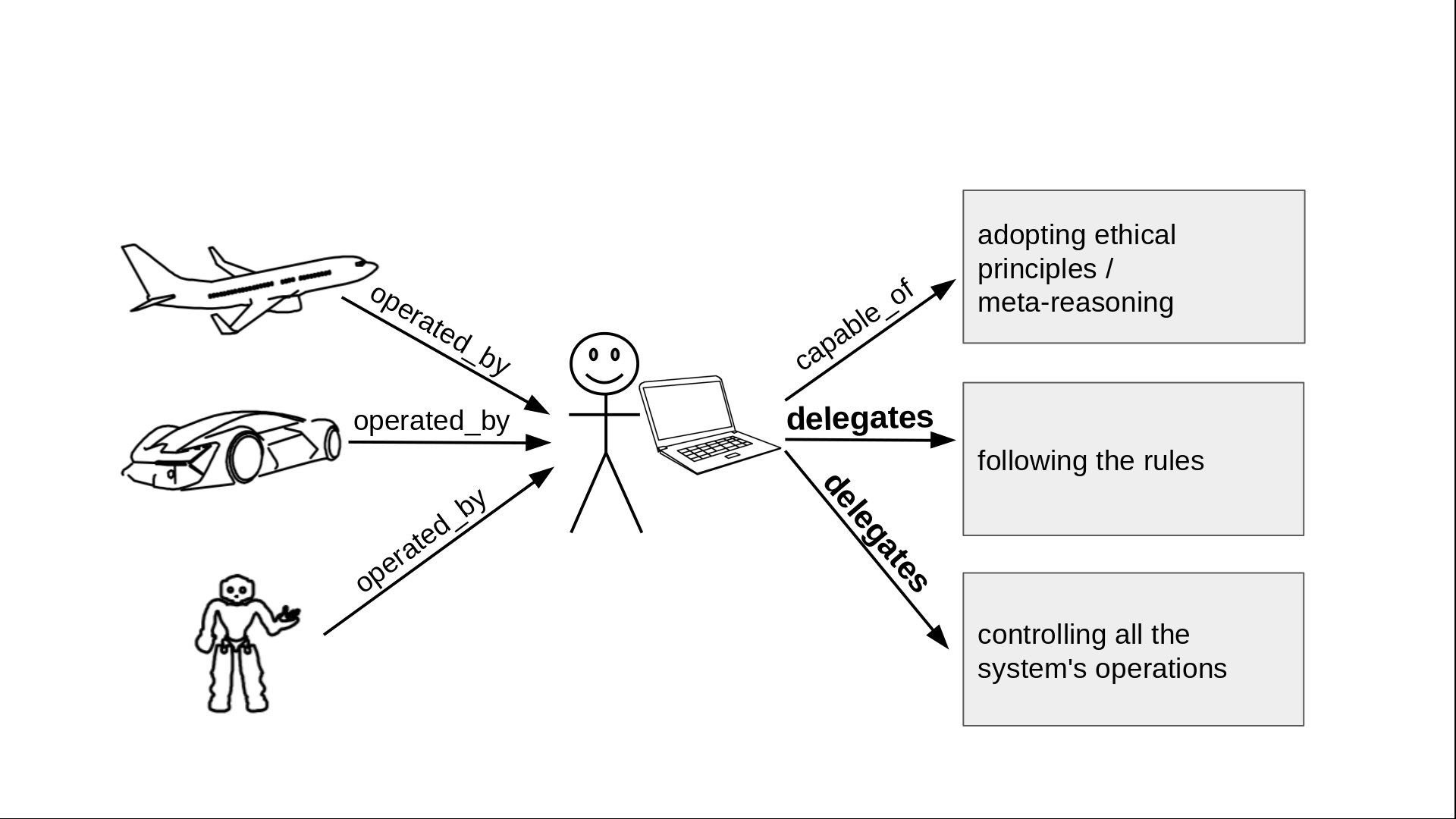}
\caption{Human (partially) delegates the respect of the rules.}
\label{fig:img3}
\end{center}
\end{figure}

Nowadays, the sophisticated systems that we rely on have
become so complex that our awareness of \emph{what} actually happens
when we exploit some of their functionality is often close to zero.
For example, how many people know how a cloud storage system works? Or
the complex link between a vehicle's brake pedal and the vehicle
speed?  Even if we are domain experts, we barely know the complete
event/data flow initiated by just pressing one button.  This is even
more true with the rise of auto-* and self-* systems (auto-pilots,
self-driving cars, self-configuring industrial equipment, \etc). We
therefore can no longer just delegate the control of basic operations.
If we want a car to drive by itself, we must also delegate to it the
requirement to follow the road traffic rules (Figure~\ref{fig:img3}).

So far, however, a self-driving car is neither designed nor
expected to make decisions in moral-ethical situations
\citep{DBLP:journals/expert/BirnbacherB17}.  When ethics, and even
merely outside-of-scope situations, bear upon autonomous operation,
the human must still be
responsible~\citep{DBLP:journals/pieee/WinfieldMPE19}.
As an example, if a self-driving car has a mechanical/software failure in a dangerous
situation or if it encounters a safety dilemma, responsibility is transferred to the human.
\begin{figure}[tb]
\begin{center}
\includegraphics[trim={2cm 1.5cm 2.6cm 3cm},clip,height=5cm]{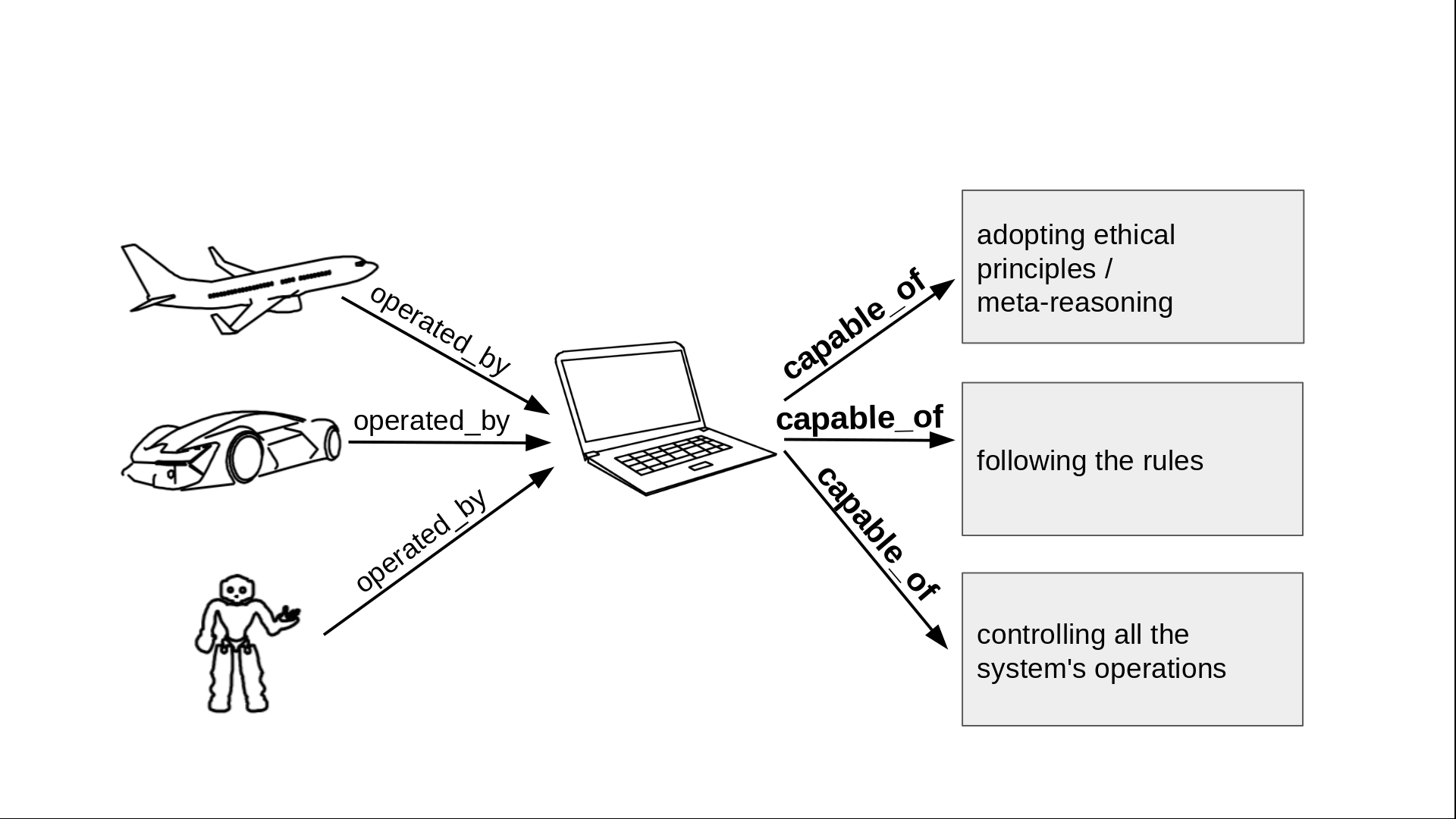}
\caption{Human replaced by an autonomous software system.}
\label{fig:img4}
\end{center}
\end{figure}

Nevertheless, due to the delegation of more and more
capabilities from humans to machines, the scenario depicted in
Figure~\ref{fig:img4} -- where the human is replaced by an autonomous
system -- is becoming more realistic.  This scenario of full autonomy
raises many ethical, legal, social, methodological and technical
issues. In this article we address the crucial question: \emph{``How
  can the reliability of such an autonomous software
  system be certified?''}

\subsection{Terminology}\label{subsec:terminology}


Before exploring this challenging question, we need to define the
terminology used in the sequel.  By `we' this article means the authors.  When we want to indicate some more general class of individuals, such as `the scientific community', or `humankind', we will explicitly use those terms.

We start with {\em reliability}. 
The term `reliable' means `suitable or fit to be relied on' \cite{reliable-definition-mw}.  For systems offering a service or
function, reliability means that the service or function is available
when needed.  A software system is reliable to the extent that it
meets its requirements consistently, namely that it makes good decisions
in all situations. 
%
In some situations, a good decision is simply
one that follows given rules, for instance, choosing to stop at a red
traffic light.  However, in other, hopefully rare situations, rules may
need to be overridden, for instance, temporarily driving on the wrong
side of the road to avoid a crash. 

Answering the question of what a `good' decision is out of the scope of this paper. Ethical decision making has been widely studied by psychologists and philosophers such as Lawrence Kohlberg 
who developed the theory of stages of moral development \cite{Kohlberg1,kohlberg1981essays,kohlberg1984essays}, and different cultures have a different attitude towards the notion of `good' decision. Our contribution is not on the philosophical challenges raised by the question, but on the technological ones.

Reliability is often associated with the notion of a \defn{certification}, `a proof or a document proving that someone is
qualified for a particular job, or that something is of good quality' \cite{certification-definition-ca}; besides the document,
certification also refers to `the process of giving official or legal
approval to a person, company, product, \etc., that has reached a
particular standard' \cite{certification-definition-be}. Human
professionals can be certified, and the idea is not new: guilds
of arts and crafts were born in the 12th century in many European
cities, to regulate and protect the activities of those belonging to
the same professional category \citep{Jovinelly06:guilds}.  Being part
of a guild was a certification of the craftman's or merchant's
professional skills. 
As soon as machines partially or completely supplemented professionals, the need to certify machines arose -- at least in terms of safety, if not functionality; this is also true of software. 
Certification of software
reliability is a lively research area in software engineering, as will
be discussed in Section~\ref{subsec:certification}.

We define a system to be \defn{autonomous} if it can make its own
decisions and act on them, without external (human) supervision and
control.  For example, a mobile robot can be completely
remote-controlled, in which case it is not autonomous, or it can have
a built-in control unit that decides on its moves, such that it
becomes \emph{semi-autonomous}.  Of course, the boundary separating
fully autonomous from non-autonomous systems is not black and white. For
example, the robot may be allowed some degree of autonomy, \eg in path
planning, whereas the overall movement goal is imposed by some remote
controller. 

\later{Perhaps also highlight LOW HIGH etc. in formatting - define macros 
}

\begin{definition}
\label{Def:Level}
The levels of autonomy that we will use to classify examples of systems from different domains in Section \ref{sec:cases},
roughly follow the six-grade scale given for autonomous road vehicles by SAE International~\cite{SAE-J3016201806-standard}, though, e.g.,  that standard does not include our `low' layer:
\begin{itemize}
\item
{\bf No autonomy:} The operator is responsible for all tasks.
\item 
{\bf Low autonomy:} Straightforward (but non-trivial) tasks are done entirely autonomously (no human poised to take over operation). 
\item
{\bf Assistance systems:} The operator is assisted by automated systems, but either remains in control to some extent or must be ready to take back control at any time.
\item
{\bf Partial autonomy:} The automated system takes full control of the system, but the operator must remain engaged, monitor the operation and be prepared to intervene immediately.
\item
{\bf Conditional autonomy:} The automated system has full control of the operation during specified tasks; the operator can safely turn their attention away but must still be prepared to intervene upon request.
\item
{\bf High autonomy:} The automated system is capable of performing all planned functions under certain circumstances (e.g., within a certain area); the operator may safely leave the system alone.
\item
{\bf Full autonomy:} The system can perform all its intended tasks on its own, no human intervention is required at any time.
\end{itemize}
\end{definition}

In addition to defining the \emph{level} of autonomy, we also consider the \emph{scope} of autonomy. This is the level of functionality of the system's autonomous capabilities. For example, one vacuum cleaner might have autonomous capabilities that only encompass traversing a space and avoiding obstacles, while another, more sophisticated model, may also be able to schedule its cleaning to avoid disruption to the human's schedule. We would say that the second model has greater scope of autonomy. The scope and level of autonomy can sometimes be a tradeoff: increasing the scope may involve the system doing things that it cannot do fully autonomously, whereas a system with more limited scope may be able to have higher autonomy. 

%


We are particularly interested in fully autonomous systems that can also make their own decisions on \emph{safety-critical} actions, namely actions whose failure could result in loss of life, significant property damage or damage to the environment.\footnote{Adapted from the definition of `Safety Critical System' provided by J. C. Knight in \cite{DBLP:conf/icse/Knight02}.}  
Additionally, autonomous systems are often characterised by the need to balance pursuing objectives over a long time period (being \emph{proactive}), 
with responding to environmental and
system changes (being \emph{reactive}).

In the sequel, we will also make strong use of the notion of an
`agent'. An \defn{autonomous software agent} (`agent' for short) is an
autonomous software system that captures the ability to decide or act
independently, while also balancing between being proactive and
reactive.  We follow standard usage in the field in defining a
\emph{multiagent system} as a system that includes multiple such
agents, which may interact in various ways (\eg communicating using
messages or via the environment): see the seminal works by Jennings N.~R., Sycara K.~P., and Wooldridge M.
\citep{DBLP:journals/aamas/JenningsSW98,DBLP:conf/ecaiw/1994,DBLP:journals/ker/WooldridgeJ95}. Finally,
we consider \emph{rational agents} as those that are structured in
terms of \emph{intentional} concepts, such as goals, beliefs, and
intentions (synonymously, the terms `cognitive agent' or `intelligent agent' are
sometimes used in the literature
\citep{DBLP:journals/ker/WooldridgeJ95}).

\begin{figure}[htb]
\begin{center}
\includegraphics[width=0.65\textwidth,clip]{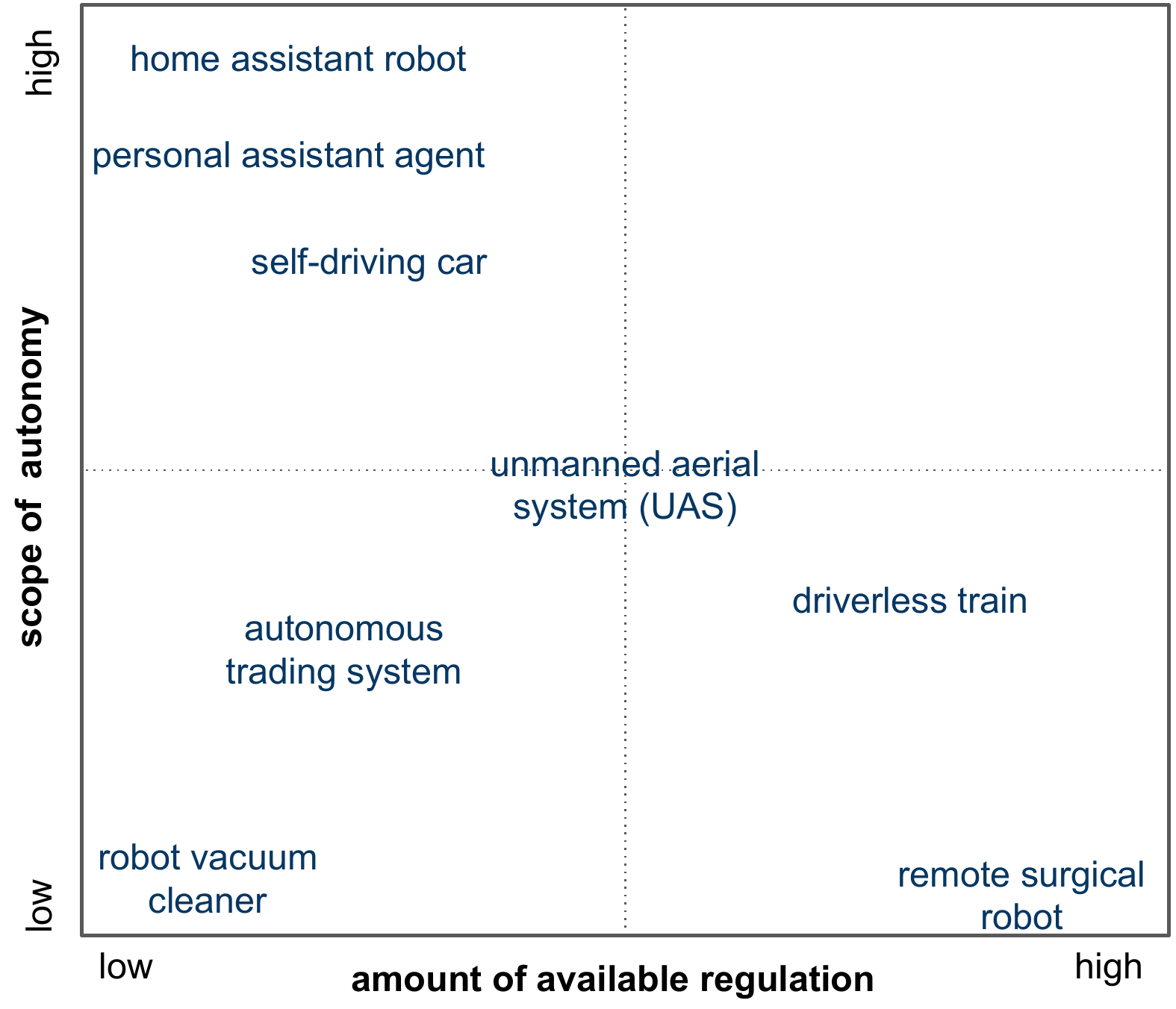}
\caption{Comparing domains of autonomous systems in terms of the level of autonomy expected in the (near) future, and the amount of regulation existing today.
Note that the level of (expected) autonomy and of existing regulation cannot be precisely measured on a scale.  This figure should be interpreted as being roughly indicative.
}
\label{fig:matrix}
\end{center}
\end{figure}

Figure~\ref{fig:matrix} compares different domains of autonomous
systems in terms of the expected autonomy and available regulation.
Although the scope\footnote{We use the scope of autonomy in this figure, rather than the level of autonomy,  because for the systems considered there is a tradeoff: the systems vary in the scope of autonomy, but for many of these systems the scope is set (by designers) in order to allow the system to display high levels of autonomy, making scope of autonomy a more useful differentiator than the level of autonomy.}
of (expected) autonomy and the level of regulation cannot be measured precisely, the figure highlights that there are systems (top left quadrant) with considerable scope for autonomy, but limited available regulation.  These are the sorts of systems that particularly require work to be able to shift them across to the right by increasing the available regulation.
We discuss each of these domains in Section~\ref{sec:cases}, with the exception of remote surgical robots, since there is not enough autonomy permitted to such systems.
It is worth noting that although many systems can be viewed as being
rational agents, we only do so when there is benefit in adopting an
intentional stance and viewing the system in these terms. For example,
a thermostat makes decisions, and we could ascribe it a goal to keep
the room at a fixed temperature.  However, the behaviour of a
thermostat is simple enough that there is no benefit to viewing it in
terms of goals and beliefs \citep{dennett1989intentional}.


It is important to highlight that, for purposes of certification, or other regulatory procedures,
we sometimes need to consider not just \emph{what} a software system did, but also \emph{why} it did it. For instance, there is a difference between a car breaking the speed limit because it has an incorrect belief about the speed limit, and a car going too fast because it believes that temporarily speeding is the best, or even the only, way to avoid an accident.\footnote{Here we see the interplay between norms of different types \citep{DBLP:journals/ker/DastaniTY18}: current jurisdiction in Germany at the time of writing is that one is not allowed to transgress speed limits even in life-threatening situations.  The argument is that even in order to save a human life one is not supposed to endanger another one.}


\subsection{Audience, Contributions and Structure}

This article assesses what is needed in order to provide verified
reliable behaviour of an autonomous system, analyses what can be done
as the state of the art in automated verification, and proposes a
roadmap towards developing certification and broader regulation guidelines.

This article thus has three audiences.  Firstly, we address regulators, who might
find the proposed roadmap useful as a path towards being able to
meaningfully regulate these sorts of systems.  Secondly, engineers and
developers who develop such systems might find it useful in
seeing how/where these systems need greater analysis.  Thirdly,
academic researchers can advance the state of the art by finding
better ways of dealing with the challenges that we articulate.


We advance the literature by
\begin{enumerate}
\item proposing a framework for viewing (and indeed building) autonomous systems in terms of three layers;
\item showing that this framework is general, by illustrating its application to a range of systems, in a range of domains;
\item discussing how certification/regulation might be achieved,
  breaking it down by the three layers; and
\item articulating a range of challenges and future work, including challenges to regulators, to researchers, and to developers.
\end{enumerate}

The remainder of the article is structured as follows.  Section~\ref{sec:where-we-are-now} reviews the current situation in terms of regulation and certification of (semi-)autonomous systems, and of the issues still open.  
Section~\ref{sec:cando} assesses what could be done in the near future; it  develops our three-layer reference framework, discusses what we need from regulators, proposes a process for deriving verification properties, and reviews in more detail verificaton techniques.  
Section~\ref{sec:cases} discusses a set of case studies in different application domains.  
Section~\ref{sec:challenges} looks at challenges in research, engineering and regulation. 
Section~\ref{sec:conc} summarises and indicates future directions.


\section{Looking Back}
\label{sec:where-we-are-now}

All systems, be they autonomous or not, that operate in a human society need to conform to some legal requirements.  These legal requirements may be generic and apply to all products, or specific.  Often these requirements are based on \defn{regulations}, that we define as `rules, policies and laws set out by some acknowledged authority to ensure the safe design and operation of systems'.\footnote{The definition provided by the Cambridge Dictionary is `the rules or systems that are used by a person or organisation to control an activity or process' \cite{regulation-definition}. We customise this definition for systems that may perform safety-critical actions.}

Relating to the concept of regulation, in the context of this paper \emph{certification} can be specified as `the determination by an independent body that checks whether the systems are in conformity or compliant with the above regulations'.
Certification involves a legal, rather than scientific, assessment and usually appeals to \emph{external review}, typically by some \defn{regulator}.

The certification
processes, and hence regulators, in turn appeal to \defn{standards}, namely documents (usually produced by a panel of experts) providing
guidance on the proving of compliance.

There are a plethora of different standards, issued by a wide range of
different standardisation organisations. Amongst the most well known are \emph{CENELEC} \cite{CENELEC-org},   
\emph{IEC} \cite{IEC-org}, \emph{IEEE} \cite{IEEE-org} and \emph{ISO} \cite{ISO-org}, to name just a few.
Many of these organisations
provide generic standards relevant across many (autonomous) system domains.
For particular
sectors, the regulatory bodies -- and there may be several for each
sector -- have a range of specific standards. 
In Section~\ref{subsec:certification-and-standards} we
present some of the most relevant existing standards, and in
Section~\ref{subsec:certification} we overview some methods and tools
suitable for certification of software systems.  
It is important to note, however,
that nowadays moving from written standards to formal specifications that can be fed to tools able to check, verify, and certify the system's behaviour, is not possible. Also, most existing standards say little, if anything, about
`autonomy' and `uncertainty', the situation where autonomy is more needed, but also more dangerous. Nevertheless, they prescribe important properties that
systems should aim to comply with. Section~\ref{subsec:openissues} faces some issues raised by autonomous systems, which are not (yet) satisfactorily addressed by current standards and regulations, including how we might link together the achievements described in the first two sections, and how we might deal with autonomy and uncertainty. 

\subsection{Standards}
\label{subsec:certification-and-standards}

Tables~\ref{tab:standards_robotics} to~\ref{tab:standards_aerospace} present
some standards grouped by domains where autonomy potentially plays a crucial
role. The most sophisticated illustrative examples in Section \ref{sec:cases}
are taken from these domains.  We do not claim this to be either exhaustive or
systematic: this section is only meant to give the reader an idea of the
complexity and wide variety of existing standards by providing examples issued
by different organisations. It is important to note that there is a \emph{vast}
array of standards, many of which are never used by any regulator.

Table \ref{tab:standards_robotics} illustrates some standards in the robotics domain. Most of them come from ISO. A Technical Committee of the ISO created in 2015 \cite{ISO-robotics-TC} is in charge of the standardisation of different robotics fields, excluding toys and military applications. In 2015, IEEE developed an ontology for agreeing on a shared terminology in robotics, and delivered it as a standard. 

\begin{table}[tb]
\caption{Examples of standards for robotics.}
{\scriptsize{
\begin{tabular}{p{1.1cm}p{3.2cm}p{0.6cm}p{7.9cm}}
\toprule
{\bf Code} & {\bf Title} & {\bf Year} & {\bf Abstract} \\
\midrule
{\bf ISO 13482} \cite{ISO-13482-standard} & Robots and robotic devices -- Safety requirements for personal care robots & 2014 & Requirements and guidelines for the inherently safe design, protective measures, and information for use of personal care robots, in particular mobile servant robot; physical assistant robot; person carrier robot.
  \\
\midrule
{\bf IEEE 1872} \cite{IEEE-1872} &  IEEE Standard Ontologies for Robotics and Automation & 2015 & A core ontology that specifies the main, most general concepts, relations, and axioms of robotics and automation, intended as a reference for knowledge representation and reasoning in robots.   
\\
\midrule
{\bf ISO/TS 15066} \cite{ISO-15066-standard} &  Robots and robotic devices -- Collaborative robots & 2016 & Safety requirements for collaborative industrial robot systems and the work environment, supplementing the requirements and guidance on collaborative industrial robot operation given in ISO 10218-1 and ISO 10218-2. \\
\midrule
{\bf ISO/TR 20218-1, ISO/TR 20218-2} \cite{ISO-202181-standard,ISO-202182-standard} &  Robotics -- Safety design for industrial robot systems -- Part 1 (End-effectors) \& Part 2 (Manual load/unload stations) & 2017, 2018 & Applicable to robot systems for manual load/unload applications in which a hazard zone is safeguarded by preventing access to it, and both access restrictions to hazard zones and ergonomically suitable work places must be considered. Guidance on safety measures for the design and integration of end-effectors used for robot systems.
\\
\midrule
{\bf ISO/TR 23482-2} \cite{ISO-234822-standard} &  Robotics -- Application of ISO 13482 -- Part 2: Application guidelines & 2019 & Guidance on the use of ISO 13482 to facilitate the design of personal care robots in conformity with ISO 13482, including new terms and safety requirements introduced to allow close human-robot interaction and human-robot contact in personal care robot applications.   
\\
\bottomrule
\end{tabular}
}}
\label{tab:standards_robotics}
\end{table}

Table~\ref{tab:standards_medical} summarises some facts of one IEC standard dealing with medical equipment. Many standards in this domain exist, also delivered by ISO which issued more than 400 standards focusing on health \cite{ISO-health} thanks to three Technical Committees dealing with medical equipment \cite{ISO-blood-TC,ISO-194-TC,ISO-210-TC} and one dealing with health informatics \cite{ISO-215-TC}. We selected \cite{IEC-TR-60601-4-1} as an example from the medical technologies domain, because it focusses on equipments with `a degree of autonomy'.

\begin{table}[tb]
\caption{Examples of standards for medical-assistive technologies}
{\scriptsize{
\begin{tabular}{p{1cm}p{3.2cm}p{0.6cm}p{8cm}}
\toprule
{\bf Code} & {\bf Title} & {\bf Year} & {\bf Abstract}  \\
\midrule
{\bf IEC/TR 60601-4-1} \cite{IEC-TR-60601-4-1} & Medical electrical equipment -- Part 4-1: Guidance and interpretation & 2017 &  Guidance to a detailed risk management and usability engineering processes for medical electrical equipment (MEE) or a medical electrical system (MES), employing a degree of autonomy (DOA) \& guidance on considerations of basic safety and essential performance for an MEE and MES with a DOA.
 \\
\bottomrule
\end{tabular}
}}
\label{tab:standards_medical}
\end{table}

Nearly 900 ISO standards have been developed for the automotive sector \cite{ISO-road-vehicles}. One of the influential is the ISO 26262 \cite{ISO-26262-1-standard}, born as an adaptation of the Functional Safety standard IEC 61508 for Automotive Electric/Electronic Systems \cite{IEC-61508-standard}. Published in 12 individual parts, ISO 26262 has been updated in 2018 to keep abreast of today's new and rapidly evolving technologies, and be relevant to even more applications. IEEE is also developing standards in the automotive sector, ranging from public safety in transportation-related events \cite{IEEE-1512-2006-standard} to system image quality \cite{IEEE-P2020-standard}. More than three dozen IEC technical committees and subcommittees  cover the standardisation of equipment used in and related to road vehicles as well as of other associated issues. As an example, the IEC TC 69 \cite{IEC-69-TC} is preparing international standards for road vehicles, totally or partly electrically propelled from self-contained power sources, and for electric industrial trucks. Table \ref{tab:standards_automotive} presents one standard for each of the three organisations above, ISO, IEEE, and IEC. 
\begin{table}[]
\caption{Examples of standards in the automotive domain}
{\scriptsize{
\begin{tabular}{p{1cm}p{3.2cm}p{0.6cm}p{8cm}}
\toprule
{\bf Code} & {\bf Title} & {\bf Year} & {\bf Abstract}  \\
\midrule
{\bf IEEE-P2020} \cite{IEEE-P2020-standard} & Standard for Automotive System Image Quality & 2016 & This standard addresses the fundamental attributes that contribute to image and quality for automotive Advanced Driver Assistance Systems applications, as well as identifying existing metrics and other useful information relating to these attributes.  \\
\midrule
{\bf ISO 26262} \cite{ISO-26262-1-standard} & Road vehicles -- Functional safety  & 2018 & 
Safety is one of the key issues in the development of road vehicles. 
With the trend of increasing technological complexity, software content and mechatronic implementation, there are increasing risks from systematic failures and random hardware failures, these being considered within the scope of functional safety. The ISO 26262 series of standards includes guidance to mitigate these risks by providing appropriate requirements and processes.  \\
\midrule
{\bf IEC 63243 ED1} \cite{IEC-63243-standard} & Interoperability and safety of dynamic wireless power transfer (WPT) for electric vehicles & 2019 & The draft of this standard, develped by the IEC TC 69, will be circulated at the end of 2019. It will specify definition and conditions of interoperability and safety for magnetic-field dynamic WPT for electric vehicles and the associated safety requirements.  \\
\bottomrule
\end{tabular}
}}
\label{tab:standards_automotive}
\end{table}

Compared to other domains, railway homologation and operation is strictly regulated. The IEC Technical Committee 9 \cite{IEC-9-TC}  is responsible for the international standardisation of the electrical equipment and systems used in railways. The ISO Technical Committee 269  \cite{ISO-269-TC} complements IEC TC 9 by addressing the standardisation of all systems, products and services specifically related to the railway sector, not already covered by IEC TC 9.
Both work in close relationship with the International Union of Railways (UIC, \cite{UIC-org}) and the International Association of Public Transport (UITP, \cite{UITP-org}). 
%
Through the CENELEC 50128 standard \cite{cenelec50128}, CENELEC assesses the conformity of software for use in railway control which may have impact on safety, \ie software whose failures can affect safety functions. 
Table~\ref{tab:standards_rail} exemplifies standards in the railway sector by presenting one standard from ISO dealing with project management, one series from IEC dealing with reliability, availability, maintainability and safety, and the CENELEC 50128 standard.
\begin{table}[]
\caption{Examples of standards in the railway domain}
{\scriptsize{
\begin{tabular}{p{1.2cm}p{3.2cm}p{0.6cm}p{7.8cm}}
\toprule
{\bf Code} & {\bf Title} & {\bf Year} & {\bf Abstract}  \\
\midrule
{\bf IEC 62278 series} \cite{IEC-622781-standard,IEC-622783-standard,IEC-622784-standard} & Railway applications -- Specification and demonstration of reliability, availability, maintainability and safety (RAMS)  & 2002, 2010, 2016
& The documents under the IEC 62278 umbrella provide Railway Authorities and railway support industry with a process which will enable the implementation of a consistent approach to the management of reliability, availability, maintainability and safety (RAMS). The process can be applied systematically by a Railway Authority and railway support industry, throughout all phases of the life cycle of a railway application, to develop railway specific RAMS requirements and to achieve compliance with these requirements.  \\   
\midrule
{\bf CENELEC 50128} \cite{cenelec50128} & Railway applications -- Communication, signalling and processing systems -- Software for railway control and protection systems & 2011   
& Specification of the process and technical requirements for the development of software for programmable electronic systems for use in railway control and protection applications, aimed at use in any area where there are safety implications. \\   
\midrule
{\bf ISO/TR 21245} \cite{ISO-21245-standard} & Railway applications -- Railway project planning process -- Guidance on railway project planning & 2018   
& Guidance on railway project planning for decision making, based upon the principles of ISO 21500 \cite{ISO-21500-standard}, by incorporating characteristics specific to railway projects.  The document is meant to be used by any type of organisation and be applied to any type of railway project, irrespective of its complexity, size, duration. It provides neither detailed requirements nor specific processes for certification. \\   
\bottomrule
\end{tabular}
}}
\label{tab:standards_rail}
\end{table}

The quantity of existing standards in the aerospace domain is huge. Established in 1947, 
ISO/TC 20 \cite{ISO-space-TC} is one of the oldest and  
most prolific ISO technical 
committees.
IEEE has published nearly 60 standards dealing with aerospace electronics, and IEC has two Technical Committees dealing with avionics-related issues \cite{IEC-97-TC,IEC-107-TC}: these committees developed about 30 standards. 
Other relevant standards bodies must be mentioned as well. The mission of the European Union Aviation Safety Agency (EASA, \cite{EASA-org}) is to ensure the highest common level of safety protection for EU citizens and of environmental protection; to provide a single regulatory and certification process among Member States; to facilitate the internal aviation single market and create a level playing field; and to work with other international aviation organisations and regulators.
The US Federal Aviation Administration (FAA, \cite{FAA-org}) summarises its mission as `to provide the safest, most efficient aerospace system in the world'. Finally, the US Radio Technical Commission for Aeronautics (RTCA,  \cite{RTCA-org}) aims at being `the premier Public-Private Partnership venue for developing consensus among diverse and competing interests on resolutions critical to aviation modernisation issues in an increasingly global enterprise'.
In Table \ref{tab:standards_aerospace} we present standards from EASA, FAA, and RTCA, including two standards dealing with Unmanned
Aircraft Systems and drones. 
\begin{table}[]
\caption{Examples of standards for the aerospace sector}
{\scriptsize{
\begin{tabular}{p{1.5cm}p{3.2cm}p{0.6cm}p{7.5cm}}
\toprule
{\bf Code} & {\bf Title} & {\bf Year} & {\bf Abstract}  \\
\midrule
{\bf RTCA DO-254} \cite{DO-254} &  Design Assurance Guidance for Airborne   Electronic Hardware & 2000 & This document is intended to    help aircraft manufacturers and the suppliers of aircraft electronic systems assure that electronic airborne equipment safely performs its  intended function.
 The
 document also  characterises the objective of the design life cycle processes and offers a   means of complying with certification requirements.\\
\midrule
{\bf RTCA DO-333} \cite{DO-333} &  Formal Methods Supplement to {DO-178C} and {DO-278A} & 2011 & Additions, modifications and substitutions to DO-178C  (see below) and DO-278A \cite{DO-278A} objectives when formal methods are used as part of a software life cycle, and the additional guidance required. It discusses those aspects of airworthiness certification that pertain to  the production of software, using formal methods for systems approved using DO-178C.\\
\midrule
{\bf RTCA DO-178B, DO-178C/ED-12C} \cite{DO-178B,DO-178C} & Software Considerations in Airborne Systems and Equipment Certification & 2012 & Recommendations for the production of software for airborne systems and equipment that performs its intended function with a level  of  confidence  in safety that complies with airworthiness requirements. Compliance with the objectives of DO-178C is the primary means of obtaining approval of software used in civil aviation products.\\
\midrule
{\bf FAA Part 107} \cite{Part107} &  Operation and Certification of Small Unmanned Aircraft Systems & 2016 & Addition of a new part 107 to Title 14 Code of Federal Regulations \cite{DOT-FAA-AR-04-36} 
to allow for routine civil operation of small Unmanned Aircraft Systems (UAS) in the National Airspace System and to provide safety rules for 
those operations. 
The rule limits small UAS to daylight and
 civil twilight operations with appropriate collision lighting,
 confined areas of operation, and visual-line-of-sight operations.\\
\midrule
{\bf  Regulation (EU) 2018/1139} \cite{EU2018-1139} &  Regulation (EU) 2018/1139 of the European Parliament and of the Council of 4 July 2018 & 2018 & First EU-wide regulations for civil drones
with a strong focus on the particular risk of the operations.
The regulations take into account the expertise of many international players in the drone domain; they will allow remotely piloted aircraft to fly safely in European airspace and bring legal certainty for this rapidly expanding industry.\\
\bottomrule
\end{tabular}
}}
\label{tab:standards_aerospace}
\end{table}
\medskip

\noindent Having reviewed relevant standards in various domains, we next turn to briefly reviewing techniques for certification of software systems.


\subsection{Certification of Traditional Software Systems}
\label{subsec:certification}

In the late 1980s, with software applications becoming more and more
pervasive and safety-critical, many scientists began to address the
problem of certifying them.  One of the first papers in this research
strand was `Certifying the Reliability of Software'
\citep{DBLP:journals/tse/CurritDM86}.  It proposed a certification
procedure consisting of executable product increments, representative
statistical testing, and a standard estimate of the mean time to
failure of the system product at the time it was released.
Subsequently, Wohlin C. and Runeson P. presented a more
mature method of certification, consisting of five steps, and suitable
for certification of both components and full systems: 1) Modelling of
software usage, 2) derivation of usage profile, 3) generation of test
cases, 4) execution of test cases and collection of failure data, and
5) certification of reliability and prediction of future reliability \citep{DBLP:journals/tse/WohlinR94}.
Further, Poore J.~H., Mills H.~D., and Mutchler D. also pointed out
that certification should be based on first generating inputs according to
the system's intended use, and then conducting statistical experiments
to analyse them. The idea that `{\em if a component has been
  verified by a mathematical proof of correctness, you may be able to
  attribute a high degree of reliability to it}' was explicitly stated there \citep{DBLP:journals/software/PooreMM93}.
This paved the way to works where the software certification of
safety-critical systems was based on formal methods. It is worth
noting that the already mentioned IEC 61508 standard
\cite{IEC-61508-standard} recommends that \emph{formal methods} be used in
software design and development in all but the lowest Safety Integrity
Levels.

Among this wide range of work, we mention those by Heitmeyer C.~L. et al.  \citep{DBLP:journals/entcs/Heitmeyer09,DBLP:journals/tse/HeitmeyerALM08}, where certification is achieved by annotating the code with preconditions and postconditions, exploiting a five-step process for establishing the property to be verified, and finally demonstrating that the code satisfies the property. In contrast to previous work by Benzel T. \citep{DBLP:conf/sp/Benzel84} and by Whitehurst R.~A. and Lunt T.~F. \citep{DBLP:conf/sp/Benzel84,DBLP:conf/csfw/WhitehurstL89} in the operating systems and database domains, Heitmeyer C.~L. et al. addressed the problem of making the verification of security-critical code affordable. 

Many mature formal and semi-formal techniques are widely used to certify software: model checking, theorem proving, static analysis, runtime verification, and software testing. While these techniques are consolidated and are indeed met while looking back, we discuss them in Section~\ref{sec33} where we present our vision of the future. The reason is that their adoption is crucial for certifying systems that are autonomous. Besides introducing them, in Section~\ref{sec33}  we compare them along the five dimensions of inputs, outputs, strengths, weaknesses, and applicability with respect to our reference three-layer framework presented in Section~\ref{sec32}.  

As observed in surveys (\eg \cite{alexander2007certification,DBLP:journals/dagstuhl-reports/CoferHHL13}),
other approaches to software certification have been exploited, besides (semi-)formal methods.
Among them, `assurance cases' were proposed as a viable way to certify safety-critical applications \cite{rinehart2017understanding}.
\begin{quote}{\em An assurance case is an organised argument that a system is acceptable for its intended use with respect to specified concerns (such as safety, security, correctness).}
\end{quote}
\citet{rinehart2017understanding}. analysed 82 works published between 1994 and 2016, concluding that transportation, energy, medicine, and military applications are the areas where assurance cases are more widely applied.

The adoption of assurance cases is rapidly spreading both in academic works and in industry: in their recent work \citep{DBLP:journals/ase/DenneyP18}, Denney E. and Pai G. present the AdvoCATE toolset for assurance case automation developed by NASA, 
and overview more than twenty research and commercial tools suitable for creating structured safety arguments using Claims-Argument-Evidence (CAE) notation \citep{CAE2010}, and/or Goal Structuring Notation (GSN) diagrams \citep{GSN2011}.


As a final remark, we observe that software certification is so challenging that the adoption of the same method or tool across different domains is often impossible. Hence, many domain-dependent proposals exist such as for robotics \citep{DBLP:conf/issre/PietrantuonoR18a}, medical systems \citep{DBLP:conf/memocode/ArcainiBGMR15}, the automotive sector \citep{DBLP:conf/issre/AreiasCIR14,yu2016automotive}, unmanned aircraft \citep{DBLP:journals/jacic/TorensAG14,DBLP:journals/jacic/WebsterCFJ14}, and railway systems \cite{DBLP:conf/rssrail/2019}. 
%
%


\subsection{Open Issues in the Certification of Autonomous Systems}
\label{subsec:openissues}

Current standards and regulations are not ready for coping with autonomous systems that may raise safety issues, and hence need to undergo a formal  process to be certified. 
One main issue in their adoption is the format in which standards are currently specified: textual descriptions in natural language. The second issue is the lack of consideration, and sometimes even of clear understanding, of the `autonomy' and `uncertainty' notions. 
Sections \ref{subsec:standards2spec} and \ref{subsec:regulatorySafety} discuss these two issues, respectively.

\subsubsection{Certifying Systems against Textual Descriptions and System Runs}
\label{subsec:standards2spec}

Let us suppose that the approach followed for the certification process is based on verification. Verifying -- either statically or dynamically -- scientific and technical requirements of complex and autonomous software applications is far from being an easy task but, at least, formalisms, methodologies and tools for representing and processing such requirements have been studied, designed and implemented for years, within the formal methods and software engineering communities.

When requirements have a legal or even ethical connotation, such as the standards discussed in Section \ref{subsec:certification-and-standards}, their verification may be harder, if not impossible.  Such `legal requirements' are written in natural language: in order to verify that a system complies with them, a step must be made to move from the natural language to a formal one.

The literature on this topic is vast, but running existing algorithms
on existing standards, and expecting to get a clean, consistent,
complete formal specification ready to be verified, is a hopeless
task.  For example, ARSENAL \cite{DBLP:conf/nfm/GhoshELLSS16} converts
natural language requirements to formal models in SAL
\cite{bensalem2000overview}, a formal language for specifying
transition systems in a compositional way, and in LTL. Although
equipped with a powerful analysis framework based on formal methods,
and despite its ability to generate a full formal model directly from
text, ARSENAL has documented limitations when dealing, for example,
with different ways to express negation and with co-locations like
`write access'.  Also, looking at the paper, the rules used to feed
ARSENAL seem to follow a simple and regular pattern, with `if' and
`when' conditions clearly defined.

Other works address similar problems in the software engineering research area \cite{DBLP:journals/software/DalpiazFFP18,DBLP:phd/dnb/Quirchmayr18,7896653}, in the agricultural regulation domain \cite{DBLP:journals/cea/Espejo-GarciaMP18}, and -- up to some extent -- in the digital forensics field \cite{yang2015information}, but the results are far from being applicable to complex, unstructured, heterogeneous standard specifications. 

Process mining \cite{DBLP:books/daglib/0027363} is an emerging discipline aimed at discovering precise and formal specifications of processes, based on data generated by instances of those processes. It builds on process model-driven approaches and data mining. There are many ways business processes can be represented using formal languages. Most of them are inspired by Petri Nets \cite{DBLP:conf/apn/Aalst02}, but there are also proposals for formalisms based on LTL \cite{DBLP:conf/bpm/MaggiMWA11}, that could be directly used to feed a model checker or a runtime monitor. 
However, in order to certify the system, the scientists in charge for the certification process would need: 
\begin{enumerate}
\item logs of real executions of the system, to mine a precise representation of its functioning (namely, a model of the system's behaviour),
\item properties that the process must verify, either represented in a logical form or translated into a logical form from a natural language description, using the techniques presented above, and
\item a model checker, for checking that the properties are verified by the system's model.
\end{enumerate}
Even if all the three items above were available, the certification would just state that the observed real executions from which the model was extracted, met the properties. Nothing can be stated on all the other not yet observed, but still possible runs of the system. The main challenge raised by this scenario is in fact that, being data-driven, the mined model only covers the already observed situations. It is an approximate specification of how a system behaves in \emph{some} normal operational scenarios meeting the rules, and in \emph{some} scenarios where rules are broken. 

At the current state of the art, certifying large and complex (autonomous) systems agains standards based on textual descriptions and system runs is out of reach, and not only because of scientific and technical obstacles: current regulations are indeed not appropriate for autonomous systems.
We note the breadth of (mainly academic) work tackling
formal methods for (autonomous) robotic systems~\cite{Luckcuck2018}
and would expect this to impact upon regulation and certification in
the future, to make them aligned with the developments in the autonomous systems area.

\subsubsection{Dealing with Autonomy and Uncertainty}
\label{subsec:regulatorySafety}
%
%
The standards and regulatory frameworks described in Section \ref{subsec:certification-and-standards} essentially
apply to existing systems, but lack some aspects we would expect of
future, more complex and autonomous systems. The first issue is
\emph{uncertainty}, the second is \emph{autonomy}. Let us
deal with each in turn.

\paragraph{Uncertainty.} Current approaches to certification and regulation often assume that
\begin{enumerate}
\item there is a \emph{finite}  set of potential hazards/failures,
\item that these can all be identified beforehand, and
\item that this finite set will not change over the lifetime of the system.
\end{enumerate}
If all the above are true then a risk/mitigation based approach can be used since we know what problems can occur.

However, as we move to much more complex environments where we cannot
predict every (safety) issue then the above assumptions become
problematic. And then, as we provide more AI components, such as
online learning modules, we are not only unsure of what the
environment will look like but also unsure of what behaviours our
system will have (since it might have learnt new ones). All these
issues pose severe problems for the current techniques for identifying
hazards/faults, assessing risk/mitigation, and building safety-cases.

In more sophisticated systems, such as a domestic robotic assistant with
healthcare and social responsibilities, improved ways of regulating such
systems will likely have to be constructed. Without such improvements, the
existing approaches will impose the above assumptions, stifling application in
all but the most static environments.

\paragraph{Autonomy.} A second issue is that the concept of `autonomy' is not well understood in existing standards/regulations. The standards mentioned so far regulate the requirements, behaviour, and development process of complex and sophisticated systems. These systems may show some degree of autonomy, but that is not their most characterising feature. The standards are neither driven nor strongly influenced by it.
Indeed, the issue of `autonomy' has been conspicuously absent from
most existing standards, as well as the ethical issues that it
raises. There are only a few, very recent exceptions.

In 2016, the British Standards Institution (BSI, \cite{BSI-org}) developed
standards on ethical aspects of robotics. The \emph{BS~8611} standard provides a guide to the \emph{Ethical Design and Application of Robots and
  Robotic Systems}~\cite{BS8611}.  
As stated in its overview:
\begin{quote}
\emph{BS 8611 gives guidelines for the identification of potential ethical harm arising from the growing number of robots and autonomous systems being used in everyday life.}

\emph{The standard also provides additional guidelines to eliminate or reduce the risks associated with these ethical hazards to an acceptable level. The standard covers safe design, protective measures and information for the design and application of robots.}

[\ldots]

\emph{The new standard builds on existing safety requirements for different types of robots, covering industrial, personal care and medical.}
\end{quote}
While the BSI feeds in to ISO standards, the above ethical standard has not yet been adopted by ISO.

In a large, international initiative, the IEEE, through its
\emph{Global Initiative on Ethics of Autonomous and Intelligent
  Systems} \cite{IEEE-ethical-initiative}, has begun to develop a
range of standards tackling autonomy, ethical issues, transparency,
data privacy, trustworthiness, etc. These standards are still in their
early stages of development; Table~\ref{tab:standards_ethics} provides
references to those which are more closely related to autonomous systems. The
year reported in the table is the Project Authorisation Request (PAR)
approval date.

\begin{table}[]
\caption{Examples of IEEE Standards related with ethics of autonomous systems}
{\scriptsize{
\begin{tabular}{p{1.1cm}p{3.4cm}p{0.8cm}p{7.5cm}}
\toprule
{\bf Code} & {\bf Title} & {\bf PAR Appr.} & {\bf Abstract}  \\
\midrule
{\bf IEEE P7000} \cite{IEEE-P7000-standard} &  Model Process for Addressing Ethical Concerns During System Design & 2016 & Process model by which engineers and technologists can address ethical consideration throughout the various stages of system initiation, analysis and design.\\
\midrule
{\bf IEEE P7001} \cite{IEEE-P7001-standard} &  Transparency of Autonomous Systems & 2016 & This standard describes measurable, testable levels of transparency, so that autonomous systems can be objectively assessed and levels of compliance determined.\\
\midrule
{\bf IEEE P7002} \cite{IEEE-P7002-standard} &  Data Privacy Process & 2016 & Requirements for a systems/software engineering process for privacy oriented considerations regarding products, services, and systems utilising employee, customer or other external user's personal data.\\
\midrule
{\bf IEEE P7003} \cite{IEEE-P7003-standard} &  Algorithmic Bias Considerations & 2017 & Specific methodologies to help users certify how they worked to address and eliminate issues of negative bias in the creation of their algorithms, 
where `negative bias' infers the usage of overly subjective or uniformed data sets or information known to be inconsistent with legislation or with instances of bias against groups not necessarily protected explicitly by legislation.\\
\midrule
{\bf IEEE P7006} \cite{IEEE-P7006-standard} &  Standard for Personal Data Artificial Intelligence (AI) Agent & 2017 &  Technical elements required to create and grant access to a personalised Artificial Intelligence (AI) that will comprise inputs, learning, ethics, rules and values controlled by individuals.\\
\midrule
{\bf IEEE P7007} \cite{IEEE-P7007-standard} &  Ontological Standard for Ethically Driven Robotics and Automation Systems & 2017 &  The standard establishes a set of ontologies with different abstraction levels that contain concepts, definitions and axioms which are necessary to establish ethically driven methodologies for the design of Robots and Automation Systems.\\
\midrule
{\bf IEEE P7008} \cite{IEEE-P7008-standard} &  Standard for Ethically Driven Nudging for Robotic, Intelligent and Autonomous Systems & 2017 &  `Nudges' as exhibited by robotic, intelligent or autonomous systems are defined as overt or hidden suggestions or manipulations designed to influence the behaviour or emotions of a user. This standard establishes a delineation of typical nudges (currently in use or that could be created). \\
\midrule
{\bf IEEE P7009} \cite{IEEE-P7009-standard} &  Standard for Fail-Safe Design of Autonomous and Semi-Autonomous Systems & 2017 &  Practical, technical baseline of specific methodologies and tools for the development, implementation, and use of effective fail-safe mechanisms in autonomous and semi-autonomous systems.  \\
\bottomrule
\end{tabular}
}}
\label{tab:standards_ethics}
\end{table}

Many efforts in the ``ethics of autonomous systems'' research field
converged in the \emph{Ethically Aligned Design} document released in
2019 \cite{EthicallyAlignedDesign2019}: the document is the result of
an open, collaborative, and consensus building approach lead by the
IEEE Global Initiative. While not proposing any rigorous standard, it
makes recommendations on how to design `ethics aware' so-called
`autonomous and intelligent systems' (A/IS), and provides reasoned
references to the IEEE P70** standards and to the literature.
  
To give an example, one of the eight general principles leading the
A/IS design is \emph{transparency} -- the basis of a particular A/IS
decision should always be discoverable.
%
The associated recommendation is as follows. 
\begin{quote}
``{\em A/IS, and especially those with embedded norms, must have a high level of
transparency, from traceability in the implementation process, mathematical
verifiability of its reasoning, to honesty in appearance-based signals, and
intelligibility of the system’s operation and decisions.}'' \cite[Page 46]{EthicallyAlignedDesign2019}
\end{quote}
While this represents a very good starting point towards agreeing on
which behaviour an A/IS should exhibit, certifying that an A/IS has a
high level of transparency, based on the recommendation above, is not
possible. Moving from well known and clear rules written in natural
language to their formal counterpart is hard, and formalising
recommendations is currently out of reach, as discussed in Section \ref{subsec:standards2spec}.


\section{Ways Forward}
\label{sec:cando}

What is the way forward?
There are a number of elements that we can bring together to
address and support regulatory development. These span across:
\begin{itemize}

\item \emph{architectural/engineering issues} --- constructing an autonomous
system in such a way that it is amenable to inspection, analysis, and regulatory
approval,

\item \emph{requirements/specification issues} --- capturing exactly how we
\textbf{want} our system to behave, and what we expect it to achieve, overcoming the difficulties arising when human-level rules do not already exist, and 

\item \emph{verification and validation issues} --- providing a wide
range of techniques, across different levels of formality, that can be used
either broadly across the system, or for specific aspects.
\end{itemize}
This second item is particularly important \cite{Roz16}: if we do not know what is expected of
the system, then how can we verify it? In traditional systems, the expected
behaviour of the human component in the overall system, be they a pilot, driver,
or operator, is often under-specified. There is an assumption that any trained
driver/pilot/operator will behave \emph{professionally}, yet this is never
spelled out in any system requirement. Then, when we move to autonomous systems,
where software takes over some or all of the human's responsibilities, the exact
behaviour expected of the software is also under-specified.
Consequently, this leads to a requirement for greater precision and level of
detail that we require from regulatory authorities and standards.

This section presents an outline for a way forward, covering the three elements.
Firstly, a key (novel) feature of the
approach proposed is a three-layer framework (Section~\ref{sec32}) that separates dealing with
rule-compliant behaviour in `normal' situations from dealing with abnormal
situations where it may be appropriate to violate rules.
For example, a system might consider driving on the wrong side of the road if there is an obstacle in its way and it is safe to use the other lane.
Secondly, we consider what we need from regulators (Section~\ref{sec31}) and define 
a process for identifying properties to be
verified by considering how humans are licensed and assessed
(Section~\ref{sec:process}).
Thirdly, we review existing verification techniques (Section~\ref{sec33}), including their strengths, weaknesses, and applicability.

\subsection{A Reference Three-Layer Autonomy Framework}
\label{sec32}

\begin{figure}[htbp]
%
  \begin{center}
        \includegraphics[width=0.7\textwidth]{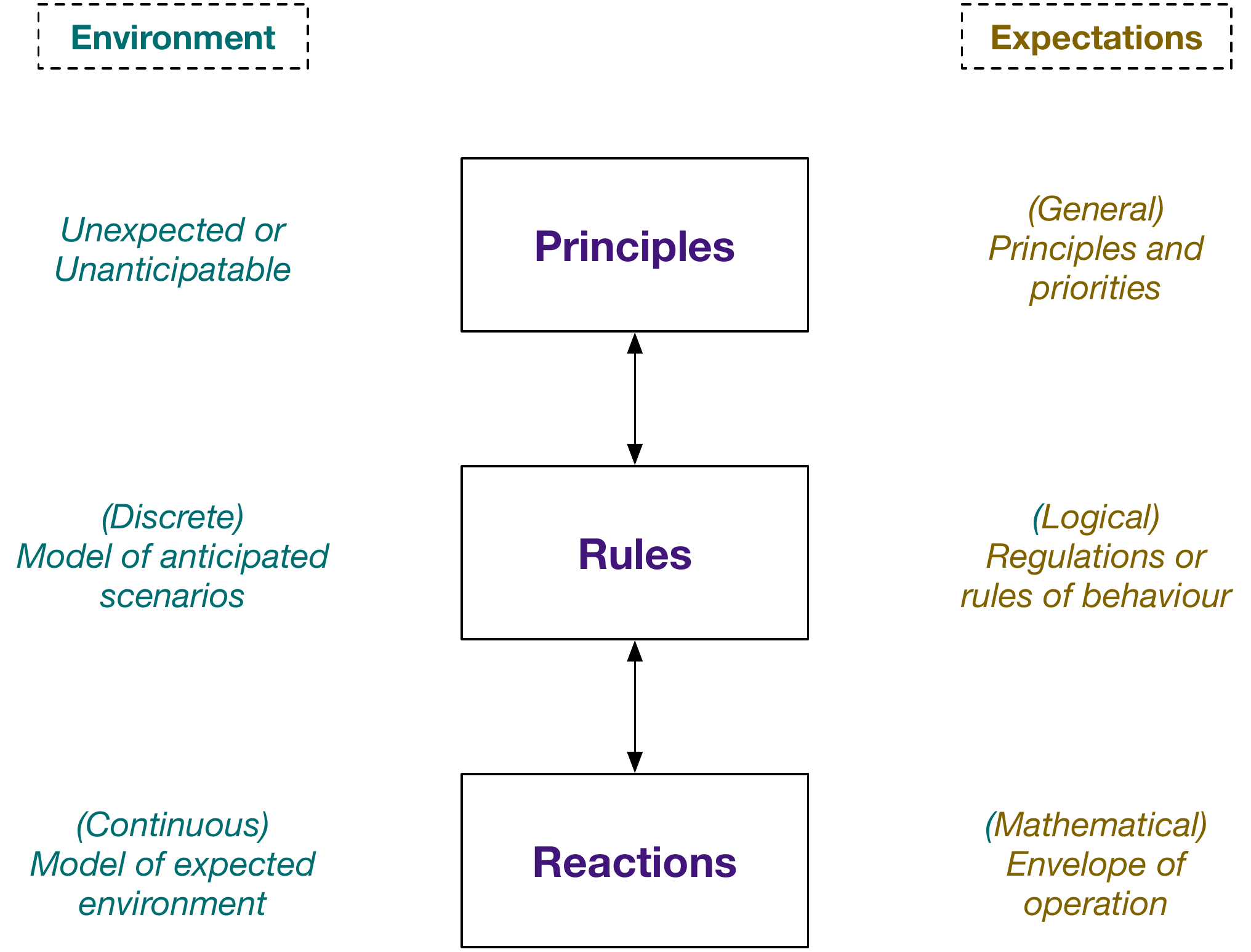}
\caption{A reference three-layer autonomy framework.}
\label{fig:three-layer-single-agent}
\end{center}
\end{figure}

In order to distinguish the types of decisions made by autonomous
systems, we present a reference three-level framework\footnote{We use `framework' rather than `architecture' for two reasons.  Firstly, to avoid confusion with an existing (but different) three layer architecture for robots. 
\later{would be nice to add a citation here}
Secondly, because this framework may not be realised in terms of a software architecture that follows the same three layers.} for autonomy
in Figure~\ref{fig:three-layer-single-agent}. This brings together
previous work on:
\begin{enumerate}
\item \emph{The separation of high-level control from low-level
  control in systems architectures.}

  This is a common trend amongst hybrid systems, especially hybrid
  control systems, whereby discrete decision/control is used to make
  large (and discrete) \emph{step changes} in the low-level
  (continuous) control schemes~\cite{FisherDW13}.

  \item \emph{The identification and separation of different forms of
    high-level control/reasoning.}

    Separate high-level control or decision making can capture a wide
    range of different reasoning aspects, most commonly
    ethics~\cite{Arkin08,BremnerDFW19} or
    safety~\cite{WoodmanWHF12}. Many of these high-level components
    give rise to governors/arbiters for assessing options or runtime
    verification schemes for dynamically monitoring whether the
    expectations are violated.

  \item \emph{The verification and validation of such architectures as
    the basis for autonomous systems analysis.}

    Fisher, Dennis, and Webster use the above structuring as the basis
    for the verification of autonomous systems~\citep{FisherDW13}. By
    separating out low-level control and high-level decision making
    diverse verification techniques can be used and
    integrated~\cite{FarrellL018}. In particular, by capturing the
    high-level reasoning component as a \emph{rational agent},
      stronger formal verification
    in terms of not just `what' and `when' the system will do
    something but `why' it chooses to do it can be carried out, hence
    addressing the core issue with autonomy~\cite{EASS:vern}.

\end{enumerate}

Our reference three-layer autonomy framework consists of:
\begin{itemize}
\item \lowlayer --- involving adaptive/reactive control/response
  aspects essentially comprised of low-level (feedback) interactions
  --- behaviour is driven by this interacting with its environment
  [e.g: `autopilot'],

\item \midlayer --- involving specific, symbolically-represented
  descriptions of required behaviours --- these behaviours are tightly
  constrained by rules [e.g: `rules of the air'],

\item \toplayer --- involving high-level, abstract, sometimes
  philosophical, principles, often with priorities between them ---
  here specific behaviour is not prescribed but principles that can be
  applied to new/unexpected situations are provided [e.g:
    `airmanship'].
\end{itemize}

\noindent We here split the high-level reasoning component further,
into \emph{rule-following} decisions and decisions based on \emph{principles} (such as ethics).
We distinguish these in that the former matches the
required rules/regulations that the system should (normally) abide by
while the latter is a layer comprising reasoning processes that are
invoked when exceptional/unanticipated situations arise (and for which
there are no prescribed regulations).

The key novelty here is the distinction between the normal operation where rules
are followed (\midlayer), and (unusual) situations where the autonomous agent
needs to reason about whether to violate rules, using, \eg ethical reasoning
(\toplayer).

\subsection{What is Needed from Regulators}\label{sec31} %


Currently, in most countries regulation responsibilities are distributed
between legislation, providing the general framework within which an autonomous
system is allowed to operate, and public authorities, which are responsible for
providing detailed rules and to supervise the conformance of systems to these
rules. In this paper, we focus on rule-making by regulatory agencies. That is,
we do not discuss legal responsibilities of the designers, producers, owners,
and operators of an autonomous system. We are concerned with behavioural
aspects of such systems, and questions arising for regulatory bodies from the
increasing autonomy.




In Section~\ref{subsec:openissues} we discussed the use of standards for verification, concluding that current approaches to certification and regulation are not adequate for verification of autonomous systems.
In this section we briefly consider what would be needed from regulators in order to allow the standards to be used to verify autonomous systems.

A key issue is that current standards are not in a form that is amenable for formalisation and assessment of software, since they are oriented solely for use by humans.
One way in which regulations are oriented towards humans, and do not
readily support regulation of software, is that regulations are framed
declaratively: a collection of statements that require substantial
(human) interpretation.  Another is that the regulations implicitly
assume, and take for granted, human capabilities and attitudes.  In
order to certify autonomous software we need the scope of
regulation to include not just low-level physical operation and sensors,
but also higher-level decision-making.  Finally, it would also be desirable for
the plethora of relevant standards to be rationalised and
consolidated.
Consequently, it may be desirable to develop separate (new) standards
for the assessment of software systems (\eg software autopilots).  
At a high level, regulations should answer the following questions.
\begin{itemize}
\item {\bf What does it mean for the system to be reliable/safe?} 
The answer to this question is a set of specifications, or the union of the following: 
  \begin{itemize}
    \item What are the regulations the system must obey? For example, the automated air traffic control system must always send a resolution to avoid two planes getting too close to each other whenever this is a possibility.
    \item What emergent behaviours are expected? For example, the automated air traffic control system should keep the airspace free of conflicts. 
    \item What would be bad? For example: the assistive robot should
      never cause harm to a human; the Therac-25\footnote{The
        Therac-25 was a computer-controlled radiation therapy machine,
        involved in at least six accidents between 1985 and 1987,
        where patients were given radiation doses hundreds of times
        greater than normal, resulting in death or serious injury
        \cite{274940}.} should never deliver radiation to a patient
      when it was not activated by hospital staff; and the automated
      air traffic control system should never instruct two planes to
      collide with each other. These are often assumptions that can be
      hard to list. They are also \emph{negative regulations}, i.e.,
      complying with these is implicit in the regulations. We need to
      explicitly state them to enable robust verification
      efforts. Certification of autonomous systems goes in two
      directions: we need to know both that the system does what we
      want \emph{and} that the system does not do what we do not want.
      This is particularly important for autonomous systems since the
      `obvious' things that a human operator would know to never do
      tend to not be explicitly captured, but can be behaviours that a
      system should (or must) avoid.
  \end{itemize}

\item {\bf How `busy' will the system be?}
The answer to this question can be in the form of
minimum/maximum throughputs, real-time bounds, or other measures. Essentially,
the specifications need some environmental context. 
For example, the automated air traffic control system may vacuously assert that it can always keep the airspace free of conflict by grounding all aircraft, or limiting the number of flights.  
However, if the specification includes an indication of the minimum level of traffic that is expected (e.g., all flight take-off requests must be granted within a reasonable time bound modulo specific exceptions), then this can prevent the autonomous system from setting such inappropriate limits or learning undesirable behaviors.
Such information, provided by regulators, might include 
bounds on how many aircraft need to be
able to fly in the airspace, the maximum allowable wait time to be cleared to
fly given safe environmental conditions, etc. 

\end{itemize}

Finally, specifications need to be compositional: they can be low-level and apply to one particular software routine or high-level and apply to the high-level architecture of the system. 
Because verification efforts are organised compositionally, as is safety cases coordination, there is a need to organise and divide the above list of types of specifications for each level/system component. 

\subsection{A Process for Identifying Requirements for Certification}
\label{sec:process}

We now present a simple process that can be used to provide guidance in identifying properties that need to be specified as verification properties for certification.
The key idea is that, if the autonomous system is performing tasks that are currently done by humans, then 
knowledge about how these humans are currently licenced can be used to help identify requirements.  So, if the humans currently
performing the task require some form of licensing (\eg driver's license, pilot's license, medical license, engineering
certification, therapy certificate, etc.), then carefully considering what the licensing process assesses and, then, how this might be assessed for an autonomous software system, would move a step towards their certification.

A motivating insight is that domains most likely to require (or benefit) from
regulation and certification of autonomous agents are those domains where humans
are very likely to have to be appropriately certified.

One challenge is that software and humans are very different in their abilities. Certain assumed characteristics of
humans, such as common sense, or self-preservation, will need to be explicitly considered and assessed for software,
even though they may not be assessed at all for humans. But even when a characteristic of the humans is assessed as part
of a licensing regime, it may well need to be assessed in a different way for an autonomous software system. For
example, a written exam to assess domain knowledge may work for humans, since limited memory requires the human to be
able to reason about the domain to answer questions, but would not work for a software system that could merely memorise
knowledge without being able to apply it.
\smallskip

\noindent We consider four key areas: 
\begin{enumerate}
\item the licensing that is used for humans;
\item the assumed human capabilities (often unassessed) that are relevant;
\item the relevant laws and regulations, and what justifiable deviations might
exist; and
\item the \emph{interface} that artefacts (\eg a cockpit) used by humans (and hence to be used by autonomous software systems replacing humans) presents.
\end{enumerate}
We now discuss these aspects in turn, beginning with licensing.

\paragraph{Licensing:}
We now consider some of the qualities that licensing might assess, and for each indicate how we might assess this quality
for an autonomous software system that is replacing a human:
\begin{itemize}
\item Physical capabilities (\eg can execute the sequence of fine-tuned adjustments required to land a plane) -- this
can be assessed for autonomous software by simulation, and assessing specific
component sub-skills. 

\item Domain knowledge (\eg does the human know all of the protocols for safe operation, how to read and interpret
domain-specific updates like Notices to Airmen (NOTAMs\footnote{A NOTAM is ``a notice distributed by means of telecommunication
containing  information  concerning  the  establishment,  condition   or   change   in   any   aeronautical   facility,   service,
procedure  or  hazard,  the  timely  knowledge  of  which  is
essential to personnel concerned with flight operations.'' \cite{NOTAM}})) -- for an autonomous software system, this would need to assess the
ability to apply (operationalise) the domain knowledge in a range of scenarios designed to require this domain knowledge
in order to behave appropriately. Note that this assessment could be in the form of a test (running some scenarios and
observing the resulting behaviour), or in the form of formal verification (showing that certain properties always hold).

\item Regulatory knowledge (\eg does the human know all of the rules, such as restrictions on flying in different
classes of airspace) -- this can be tested similarly to domain knowledge.

\item Ethical normalisation (\eg does the human understand the
  importance assigned to the hierarchy of regulations from the
  regulatory body such as the FAA). An example would be that if an Unmanned Aerial System (UAS)
  is going to crash, the remote human operator needs to understand
  that it is better to clearly communicate to the appropriate air
  traffic control authority that the UAS will violate a geofence
  surrounding an unpopulated area and proceed to do that. The
  alternative, obeying the geofence but crashing the UAS into a
  populated area, is not acceptable -- for autonomous software, one
  could verify certain priorities, if the reasoning is explicitly
  represented in a decision-making component, or assess performance in
  scenarios designed to present these sort of ethical challenges.
\end{itemize}

\paragraph{Assumed human capabilities:}
Various characteristics of humans may not be assessed at all, but simply assumed,
since they are sufficiently universal, and, for some characteristics, it is very clear if they are absent (\eg a human
lacking physical mobility (lacking requirements for \emph{physical capabilities}), or being a child (lacking requirements for advanced \emph{ethical normalization}) would be clear without requiring explicit assessment). 
Specifically, in considering assessment of autonomous software systems, we would want to carefully consider what human
characteristics are required and assumed to hold, without any assessment. Typical questions might include:
\begin{itemize}
\item 
For example, we assume a pilot knows that, when flying a passenger plane, the passengers require a certain amount of time to fasten seat belts.
\item Does the human need assumed (but untested) physical capabilities? For example, a pilot can sense when something is wrong from certain sounds or vibrations in a plane that may not be present in simulations or ground tests for certification. 
\item Does the human need a certain level of maturity or life experience? For example, a sound may be readily identifiable as `bad' even if the pilot has never heard it previously.
\item Does it assume basic properties of human values/ethics, such as that the pilot would never crash the plane on purpose because the pilot has a strong inclination toward self-preservation? Does it assume an operator would never choose to hurt others? 
\end{itemize}
On the other hand, certain human characteristics that need to be
assessed when certifying humans may not need to be considered when
certifying software. For instance, psychological state, or personality
aspects (such as being aggressive, having a big ego and therefore
being over-sensitive to criticism of flying ability, or being
impulsive) should not need to be considered for autonomous agents.



\paragraph{Legal factors:}
Often licensing
includes testing for knowledge of relevant laws and regulations. We consider legal factors separately because this is a
key source of specifications, and because the licensing assessment may not cover all relevant regulations and laws. 

As per the three layers framework, we need to identify not just the `by the book' rules (\eg regulations and laws).
Rather, we also need to consider situations where these rules may need to be (justifiably) overridden, and the
principles that would need to be used to make such decisions. The key questions are: in what situations should the rules
be overridden? How can the system identify these situations? And how can the system decide what to actually do in these
situations? More specific questions to be considered include: 
  \begin{itemize}
  \item What situations could lead to outcomes considered to be `bad', `unsafe', `insecure', or otherwise universally to-be-avoided if at all possible? 
  And how bad are each of these? Are some worse than others? If there is a choice, is there any ranking or cost (possibly probabilistic) that can be associated with each? For example, if an autonomously-operating UAS is going to crash and must choose between crashing into a pile of trash on a curb or the car parked next to the pile of trash, the cost function would be expected to steer the crash landing toward the pile of trash. This could be defined in terms of minimising the repair cost of the affected property. One might furthermore define the cost of harming a human in a crash as higher than any situation where a human is not harmed.
  \item Are some rules more important to obey than others? 
	Are there divisions of the rules into hard constraints versus soft constraints?
	Is there some partial ordering on the importance of the rules?
  \item Are there any acceptable reasons to break the rules?
  \end{itemize}
In order to develop a system that can meet the requirements, we need
to also consider what are the computational requirements of the
system. What does it need to be able to measure, deduce, or decide?
  
Note that context is often left unspecified but it importantly restricts the applicability of licensing. To take context into account, we identify the licensing requirements, and then, for each requirement, consider whether it is relevant for the system.  For example, for an auto-pilot that is only used while the plane is at
cruising altitude, we would not need to consider requirements related to landing, or interaction with ground obstacles, even though these are requied for a human to earn a pilot's license.

\paragraph{Human-system interface:}
There is also a collection of similar factors that relate to the interface that a human currently uses to interact
with artefacts. Such artefacts are designed for human use, and if a human performing the task is going to be replaced by
an autonomous software system, then whether the existing interface presented by the artefact
embodies assumptions about humans should be taken into account. Specifically:
\begin{itemize}
\item Does the interface assume physical shapes of humans, such as being operated by a human hand?
\item Does it assume physical limitations of humans, such as having a minimum reaction time or a maximum speed for selecting multiple controls in sequence?
\item Does it assume mental limitations of humans, such as taking into account that a human cannot instantly take control of a system but requires orientation to the operational context to make reasonable decisions?
\item Does it assume that human-type faults will occur, such as being designed to avoid human confusion modes?
\item Does it assume that common sense deductions on the part of the operator are automatic? For example, a human pilot will
automatically notice if 
 the wing detaches from the aircraft and there is no explicit requirement that the aircraft operator must continuously ensure the aircraft has both wings but an autonomous system would need to be explicitly designed to consider this a fault (not just the instability it causes) and, e.g., avoid future control decisions that would only work if the aircraft had two wings \cite{Tom03}. There is not a sensor for every `obvious' fault yet an autonomous system needs to account for all faults that are obvious to humans, even when they trigger fault warnings for unrelated errors.\footnote{This is not an
artificial example: there was a case of an engine exploding, resulting in a large number of seemingly unrelated alerts \cite{FAA10}.
Thankfully, in that instance, the human pilot realised what had happened, and was then able to safely land the plane.}
\end{itemize}
There are three options to deal with a situation where a human is
being replaced (partially or completely) with software, and the human
interacts with an existing artefact using an interface.  These options
are: to retain the interface, and have the software interact with it,
to extend the artefact with additional interfaces for the software, or
to replace the artefact's interface completely.
The process for identifying requirements is summarised in Figure~\ref{fig:proposal}. 

\begin{figure}
\fbox{\begin{minipage}[c]{\textwidth}
\begin{enumerate}
\item \textbf{Consider licensing requirements}
	\begin{itemize}
	\item If physical skills are assessed then use simulations to assess software's sensors and affectors
	\item If domain and/or regulation knowledge is assessed (\eg using a written test), then assess the software
	using scenarios designed to assess the operationalisation of this knowledge
	\item If ethical aspects are assessed then:
		\begin{itemize}
		\item If the software system includes explicit reasoning about these aspects, then verify these rules, 
		\item Else assess the system using scenarios designed to assess ethically-appropriate behaviour
		\end{itemize}
	\end{itemize}
	
\item \textbf{Consider (typically assumed and unassessed) human characteristics}
	\begin{itemize}
	\item Basic knowledge and capabilities (\eg maths, reading, basic folk
psychology \& physics)
	\item Basic common sense, maturity, life experience
	\item Values and ethics (\eg self-preservation)
	\end{itemize}
	
\item \textbf{Consider relevant regulation and laws}
	 including justifiable deviations from the rules
	
\item \textbf{Consider human-system interfaces}
	\begin{itemize}
	\item What assumptions does the interface make
about the operator being a human?
	\item How to replace the interface:
		\begin{itemize}
		\item retain it
		\item extend it with additional interfaces
		\item replace it
		\end{itemize}
	\end{itemize}
\end{enumerate}
\end{minipage}
}
\caption{Summary of the process for identifying requirements.}
\label{fig:proposal}
\end{figure}

\later{update of Figure 7: the text in this
figure is OK but less refined than the paper's main text surrounding it.}

\subsection{Verification of Autonomous Software Systems}\label{sec33}


In Section \ref{subsec:certification} we observed that the most solid  and reliable way to certify autonomous software systems would be the adoption of \emph{formal methods}.  

\emph{Formal methods} are mathematically rigorous techniques for design, specification, validation, and verification of a wide variety of systems. They contribute to certification in a number of ways, from requirements design, to checking that all requirements can be true of the same system at the same time (before the system is built) \cite{RV10}, to verifying that early (or even partial) designs always satisfy requirements, to generating test cases, to checking during system runtime that the system is still meeting requirements. 

In order to verify that a system operates correctly, we  need to know how the system works. If we do not have
sufficient information on both how the system operates and what it means for the
system to operate safely, then it is impossible to verify that these two
behaviour spaces match up with each other. Knowing how the system works includes
knowing sufficient implementation details to be able to certify correctness. 

Though each method works differently, intuitively all formal methods verify that the system -- or a model of the system -- does what we expect it to do and nothing else. 
This capability is required for certification, \eg via \emph{safety cases} per the standards of the aerospace industry, where certification requires making a case, backed by evidence, that the system meets standards for safety and reliability. 

In some formal methods, such as model checking and theorem proving,
both the system under certification and the properties to be verified
are modelled using a rigorous, mathematical or logical language. We
indicate these methods as \emph{formal at the property \& system
  level}. The system model itself, however, is necessarily an
abstraction of the real system and hence it is incomplete: any results
from methods that operate, albeit in a rigorous way, on abstractions
of real systems, should be validated against the actual environment. (Note that many modern model checkers and theorem provers are now capable of generating software from their proven models that can be used as a basis for, or as the entire, real system implementation thus enabling straightforward validation.) 
Other methods model the properties to check using rigorous formalisms and languages, and check the property against the \emph{real} system under certification. We define these methods, which include some static analysis approaches and runtime verification, as \emph{formal at the property level}.
%
\emph{Semi-formal methods} specify either the system or the properties that they should meet using languages with informal or semi-formal  semantics. UML \cite{Booch:1999:UML:291167} and Agent-UML \cite{DBLP:conf/aose/BauerMO00} are examples of semi-formal specification languages, as well as test cases.
\emph{Informal methods} are based on specifications provided in natural language and are out of the scope of our investigation.

In this section we review five verification methods: model checking,
theorem proving, static analysis, runtime verification, and systematic
testing. The first four of these are usually categorised as formal or semi-formal
methods. Software testing is not a formal
method, but it is one of the most widely adopted verification approaches in
industry, often associated with quality assurance.  The `Software
Testing Services Market by Product, End-users, and Geography -- Global
Forecast and Analysis 2019-2023' \cite{ST2019} foresees that the
software testing services market will grow at a compounded average
growth rate of over 12\% during the period 2019-2023.  Software
testing -- once automated and seamlessly integrated in a development
method, as in the `continuous integration model' -- 
may indeed
represent 
a useful complement to formal methods. Testing can be measurably improved by using artefacts from formal verification for test-case generation. However, as is well-known, testing is not exhaustive, which limits its utility.

\begin{figure}
\begin{center}
\includegraphics[width=0.8\textwidth]{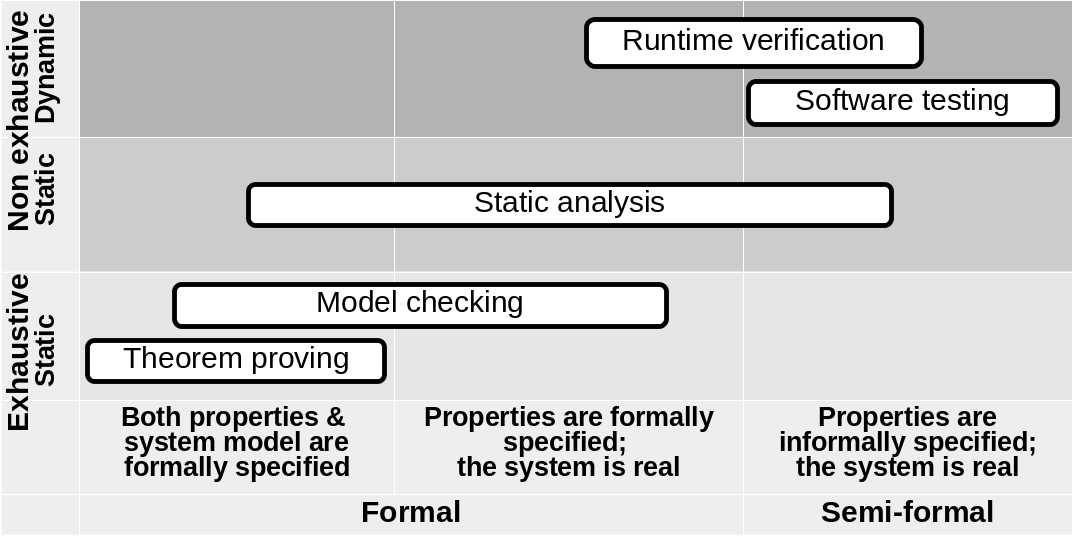}
\caption{Knowledge map of the verification methods discussed in this section.
}
\label{kmvm}
\end{center}
\end{figure}

\begin{figure}
\begin{center}
\includegraphics[width=0.5\textwidth]{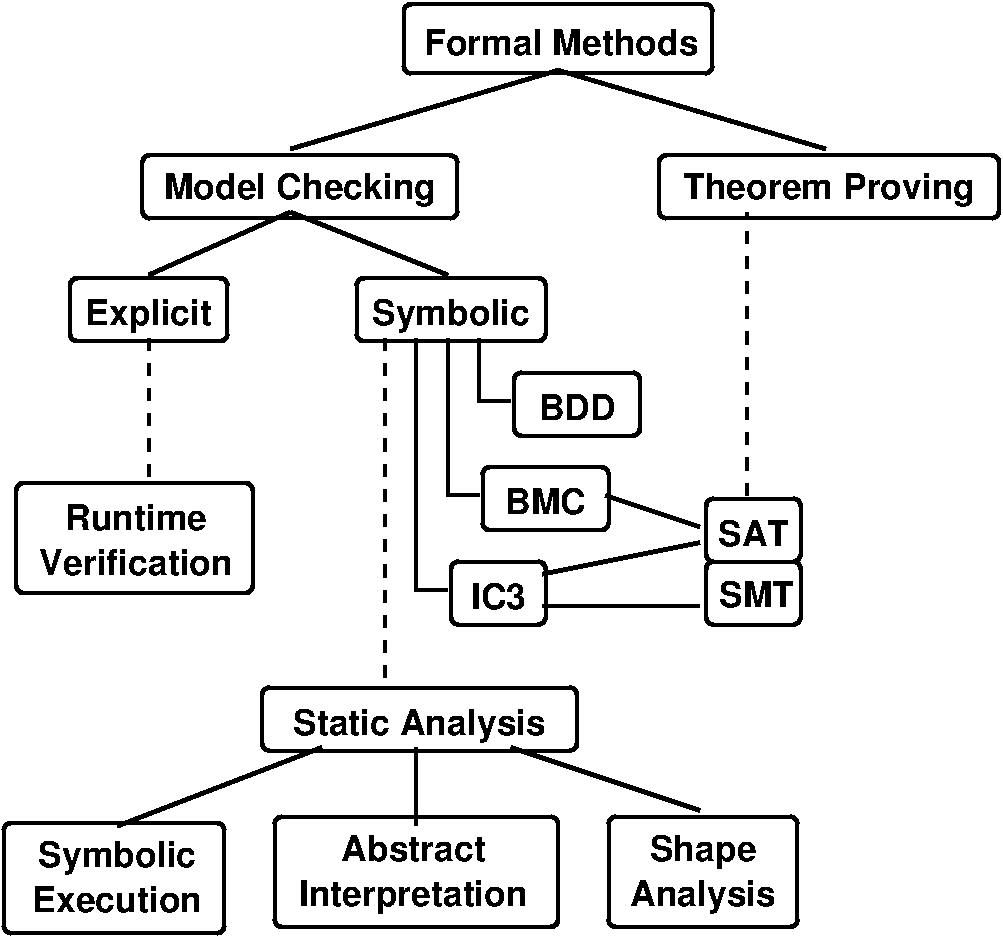}
\caption{The tree of Formal Methods; solid lines represent direct variations whereas dashed lines represent related derivations. Symbolic model checking \cite{Roz11} is divided into three major algorithms under the hood: BDD (binary decision diagram), BMC (bounded model checking), and IC3 (a pseudo-acronym sometimes replaced with PDR for property-directed reachability). As a subtlety, the BMC and IC3 algorithms are not complete by themselves but many tools use these acronyms to refer to those algorithms supplemented with proofs to make them complete. Both model checking and theorem proving commonly use SAT (Boolean satisfiability) and SMT (satisfiability modulo theories) solvers under the hood. 
}
\label{FormalMethodsTree}
\end{center}
\end{figure}

Figure \ref{kmvm} positions the five methods in a space characterised by the \emph{formality}, \emph{exhaustivennes}, and \emph{static vs.~dynamic} dimensions. Figure \ref{FormalMethodsTree} details the formal methods and their relationships. 

A method is \emph{exhaustive} when it verifies the entire behaviour space, over all possible
inputs, including for an infinite input space.  Exhaustive verification includes proving both the existence of `good' or required behaviours and the absence of `bad' or requirements-violating behaviours, something that can never be demonstrated through testing or simulation. Being exhaustive refers to the capability of exhaustively exploring the state space generated by the system model, as it is not currently possible, except for a small class of systems with specific characteristics, to be exhaustive with respect to a system of realistic industrial size. 

A method is \emph{static} when it is applied on the system's code, or on its abstract/simplified model, without needing the real system to be executed. A method is \emph{dynamic} when it operates while the system runs.

\paragraph{Model Checking:} This method performs an exhaustive check that a system, given in a logical language, always satisfies its requirements. 
Intuitively, model checking explores all possible executions of the system,
checking the system model cannot reach states that violate the provided verification properties.
\begin{itemize}
\item {\bf Inputs:} (1) a logical description of the system (either a model in a
logical language such as SMV \cite{McM99} or AIGER \cite{BHW11}, or the system itself, \eg
for software systems); (2) a requirement in a temporal logic such as LTL. See \cite{Roz11} for a survey with examples.
\item {\bf Outputs:} (1) an automated (push-button) proof (usually by assertion,
without the full proof argument) that the system always satisfies the
requirement; or (2) a counterexample, which is an execution trace of the system
from system initialisation that shows a step-by-step path to a system state
where the system has operated faithfully to its definition yet still violated
the requirement.
\item {\bf Strengths:}
\begin{itemize}
\item {\bf Automated.} Counterexamples are excellent maps for debugging and are generated without guidance from the user.
\item {\bf Exhaustive. } Verification reasons over the entire state space. 
\item {\bf Verification of Absence of Bad Behaviours.} Model checking both gives the
existence proof that a system always does what we expect given the inputs
\emph{and} that the system never does something we don't expect for any input.
In other words, given an appropriate specification, it verifies the absence of behaviours that we don't want, whether
they are triggered by an input or no input at all.  
\item {\bf Incremental.} Model checking can efficiently analyse partial models/partial systems;
can be used from the earliest stages of the design process to save time pursuing
designs that cannot satisfy their requirements based on the partial
design-so-far.
\end{itemize}
\item {\bf Weaknesses:} Model checking is garbage-in, garbage-out; the results
are only as accurate as the input system/model and the input requirement. One approach for validating the results of model checking against the actual environment by recognising assumption violations at runtime has been proposed in \cite{DBLP:conf/atal/FerrandoDA0M18,DBLP:conf/rv/FerrandoDA0M18}, but the example addressed there is simpler than any real autonomous system.
Modelling a system in a logical language and specifying a requirement in a
temporal logic are the biggest bottlenecks to the use of model checking
\cite{Roz16}; both tasks are difficult to complete and to validate. It is easy to create models that are too large to analyse with current technology, such as large pieces of software or models over large numbers of variables. Not all logical systems can be specified in languages that can be model-checked.  


\item {\bf Applicability:} model checking is widely used for both hardware and software,
   such as that seen in the \midlayer%
~\cite{BaierKatoen08:book,clarke.00.mc,ClarkeS01,Holzmann03:spinbook}. System protocals are particularly amenable to model checking; it has successfully verified real-life, full-scale communication protocols \cite{musuvathi2004model,edelkamp2004directed}, air traffic control systems \cite{ZR14,MCGTR15,GCMTR16}, wheel braking systems \cite{BCPJKPRT15}, and more. Model checking
   has been used for the \lowlayer, but quickly reduces to either
   hybrid model checking or the verification of coarse
 abstractions~\cite{AbateKM11,AlurDiscreteAbstractions00,HenzingerHW97,ApStoHySys09}. Hybrid
   model checking can involve a wide array of
   numerical techniques aimed at solving, for example, differential
   equations concerning control or environmental models. For the
\toplayer, model-checking has been used for verifying BDI agents \cite{DBLP:conf/atal/BordiniFPW03,DBLP:journals/expert/BordiniFVW04,DBLP:journals/aamas/BordiniFVW06}, 
epistemic and temporal properties of multiagent systems \cite{DBLP:conf/atal/KongL17,DBLP:conf/atal/KongL18,DBLP:conf/atal/PenczekL03}, the agents knowledge, strategies and games \cite{DBLP:conf/atal/LomuscioR06}, and general,
high-level decision making~\cite{EASS:vern} that has been extended
to ethical principles~\cite{BremnerDFW19,DennisFSW16}. 
\end{itemize}
%
%
We note one important variation, \emph{program
  model-checking}~\cite{VisserHBPL03}. Per its name, program model checking replaces the abstract model with  the actual code/software system. It uses the same model-checking algorithms, 
but on symbolic executions of the system. The advantage of such an approach is that
 it simplifies model validation (\eg proving the link between the model and the real
program/system) since we directly check the real
program. However, a disadvantage is that the use of techniques such
as \emph{symbolic execution} can make the checking process slow and
complex. Finally, note that the \toplayer{} model-checking of
high-level, and ethical, decision-making described
above~\cite{EASS:vern,DennisFSW16,BremnerDFW19} actually utilise
program model checking over rational agents~\cite{MCAPL_journal,dennis18:mcapl}.

\paragraph{Theorem Proving.} A user creates a proof that a system satisfies its requirements that is checked by a known-to-be-correct proof assistant. Some automation of standard logical deductions is included in the proof assistant. 
\begin{itemize}
\item {\bf Inputs:} A theorem asserting a requirement holds over the
  system; the same large proof includes relevant definitions of the system and (sub-)
  requirements and any necessary axioms.
\item {\bf Outputs:} A computer-checked proof that the theorem holds. If no such
proof can exist (\eg because the theorem does not hold), the user might gain the necessary insight into why this is the case from the way in which the proof fails. 
\item {\bf Strengths:} Theorem proving can provide strong type-checking and
generate proof obligations that provide unique insight into system operation.
The proofs generated are re-playable: they can be re-run by any future user at
any time. If the input theorem changes (\eg to perform a proof over the next version of the system) the old proof can be replayed (in some contexts) and only prompt the user for input where small changes are needed to update the proof to the new theorem. Proofs can be built up in extensible libraries and hierarchically inherit from/build upon each other. Large swaths of mathematics important to assertions over safety-critical systems have been proved in free libraries associated with the most popular theorem proving tools. Theorem proving can, and has, checked very large systems, even beyond what humans can inspect, such as the four-colour theorem \citep{Appel:four-colour}.
\item {\bf Weaknesses:} This formal method requires the most experience from
users and has a comparatively high learning curve relative to other forms of verification. Setting up a proof correctly is a difficult activity even for trained experts. The user is the sole source of insight into why proofs fail; there is no automated counterexample like with model checking.
\item {\bf Applicability:} Theorem-proving has been widely used for
  software seen in the \midlayer, and techniques such
  as Hoare Logic~\cite{Hoare:1969:ABC:363235.363259} and tools such as
  Isabelle~\cite{Isabelle94} are often used there. Theorem-proving has
  been used for the \lowlayer, most notably verification of NASA's aerospace systems via PVS \cite{galdino2007formal,pike2007modeling,munoz2013tcas,munoz2015daidalus}, verification of, e.g., automotive protocols through the KeyMaera system~\cite{platzer08:_keymaer,Platzer10}, and the use of theorem-proving tools to prove control
  system stability~\cite{8107009}. See \cite{rw1996introduction} for an introduction to PVS specification for a Boeing 737 autopilot. Theorem-proving for the \toplayer\ is less common, though it has been tackled~\cite{BringsjordAB06}.
\end{itemize}

\paragraph{Static Analysis.} During system design time, an automated tool examines a program or code fragment without executing it, to check for common bugs. Static analysis can be combined with techniques for expanding the types of bugs found, including symbolic execution (analysing a program's control flow using symbolic values for variables to create a map of program execution with provable constraints), abstract interpretation (creating an abstraction of the program's semantics that enables proofs about its execution), and shape analysis (mapping the program's memory signature through its data structures).
\begin{itemize}
\item {\bf Inputs:} Program or code fragment to analyse
\item {\bf Outputs:} List or visualisation of possible code defects
\item {\bf Strengths:} Automated tools analyse the structure of the code rather than just paths through the code like testing; they can therefore provide a deeper analysis with better code coverage. Static analysis is faster and more reliable than the alternative of manual code review. Though performance depends on the programming language and static analysis tool, performing static analysis is usually totally automated with easy-to-understand output suitable for a non-expert in verification.
\item {\bf Weaknesses:} Static analysers have no understanding of developer intent (e.g., they can detect that a function may access unallocated memory but will not find that the function does a different task than intended/labeled in the comments). They cannot enforce full coding standards, e.g., because some coding rules are open to interpretation, like those concerning comments, good documentation, and understandable variable names. The above ambiguities (in programmer intent, coding standards, use cases, etc.) can lead to false positive and false negative defect reports.
False negatives mean that the analyser has failed to identify an error, can lead to fallacious trust in the software.
False positives mean that the analyser reports an error where there is none, can cause a lot of wasted work at the developer's side.
\item {\bf Applicability:} 
Static analysis covers a range of different methods.
Basic checks, i.e., whether each used variable was previously initialised, are included in most modern compilers.
More sophisticated checks, e.g., whether pointer access may lead to an exception, are realised by widely used tools such as Lint \cite{lint}.
 High-end static analysers such as Polyspace \cite{polyspace} and Astr\'ee \cite{astree} are able to check for general assertions.
 A review of 40 static code analysis tools updated to November 2019 is available via the Software Testing Help blog \cite{scat}.
 According to CENELEC EN50128, static analysis is highly recommended for all safety-relevant software.
Consequently, static analysis has been applied for most safety-critical software in avionics, railway and automotive; e.g., NASA regularly releases its own home-grown static analysis tools including (Java) Symbolic PathFinder
 \cite{luckow2014symbolic}, IKOS  (Inference Kernel for Open
Static Analyzers) \cite{brat2014ikos}, and C Global Surveyor (CGS) \cite{brat2005precise}.
Static analysis techniques generally analyze software in the \midlayer. The \lowlayer\ requires analysis techniques able to deal with continuous physical models, and the \toplayer\ requires the ability to deal with potentially complex reasoning. Although static analysis can analyze autonmous systems' software, its key weakness, being non-exhaustive, limits the utility of applying it.
 \end{itemize}

\paragraph{Runtime Verification.} Runtime verification (RV) is a semi-formal method aimed at checking \emph{the
current run} of the system. RV is most often run online (\eg embedded on-board
the system during operation), and stream-based (\eg evaluating requirements
continuously throughout system operation), though it can also be run offline
(\eg with recorded data for post-mission analysis). 
\begin{itemize}
\item {\bf Inputs:} 
(1) Input data stream or streams containing time-stamped
sensor or software values, \eg sent over a system bus to the RV engine; (2) A requirement to verify, expressed in the form of a temporal logic formula, most commonly Mission-time Linear Temporal Logic (MLTL)~\cite{RRS14,LR18,LVR19}  or First-Order Linear Temporal Logic (FOLTL)~\cite{EME90,BKMP08}  whose variables are set by the input data 
or operationally via finite state automata, trace expressions \cite{DBLP:conf/birthday/AnconaFM16}, etc. (see Sect 2.1 of \cite{DBLP:conf/rv/FalconeKRT18}). 
\item {\bf Outputs:}  For online, stream-based RV, the output is a stream of
$\left< \mathrm{time}, \mathrm{verdict} \right>$ tuples containing a time stamp
and the verdict from the valuation of the temporal logic requirement (true or
false), evaluated over the data stream starting from that time step. Variations
on RV include extensions to non-Boolean (\eg Bayesian or otherwise probabilistic) results, and evaluations of only part of the data stream of interest. 
\item {\bf Strengths:} RV is the only formal verification technique that can run
during system deployment. Sanity checks can provide near-instant confirmation
that a system is obeying its requirements or that a failure has occurred,
providing an efficient mitigation trigger. Like simulation, RV can also provide useful characterisations of system runs, though RV analyzes runs in terms of violations of formal specifications; see \cite{Roz19} for a detailed comparison.  
\item {\bf Weaknesses:} Specification of the requirement to monitor at runtime
is the biggest bottleneck to successful deployment of RV \cite{Roz16}. Specifications are
difficult to write for the same reasons as for model checking; further
complications include noisiness of sensor data and challenges of real-world
environments. Limitations and constraints of embedded systems and certification processes challenge RV implementations, though to date two tools have risen to these real-world-deployment challenges and more are sure to follow \cite{RS17,AFFST17}.
Of course, violation of certain critial safety properties is not acceptable, even if this is detected. Since RV operates at (or after) runtime, it
is not suitable for such properties. 

\item {\bf Applicability:}  RV is widely used for
software in \midlayer\ as discussed for example by \citet{Rosu12} and by \citet{HavelundR17}; see \citet{TRV12} for another example. 
RV of multiagent systems using trace expressions \citep{DBLP:conf/atal/AnconaFM17,DBLP:conf/atal/FerrandoAM17}  is also a viable approach to cope with the \midlayer.
RV has been used for the \lowlayer, for
example by \citet{BartocciBBMS17}.  The R2U2 framework  \cite{RRS14,GRS14,SMR16,RS17, MRS17} is a real-time, Realisable, Responsive, Unobtrusive Unit for runtime system analysis, including security threat detection, implemented in either FPGA hardware or software. R2U2 analyzes rules and reactions spanning the \midlayer and (with the addition of a parallel system model) the \lowlayer of our reference framework.
For the \toplayer, \citet{DBLP:journals/ker/DastaniTY18} survey the use of norms for monitoring of the behaviour of autonomous agents: monitoring of norms is indeed foundational for processes of accountability, enforcement, regulation and sanctioning. 
The idea
of a runtime ethical monitor, or governor, is also commonly used~\citep{Arkin08,WoodmanWHF12}.
\end{itemize}

\paragraph{Software Testing.} Software Testing (ST) is not a formal method, though formal methods are often used for test-case geeration with provable qualities, such as coverage metrics. ST amounts to observing the execution of a software system to validate whether it behaves as intended and identify potential malfunctions  \cite{DBLP:conf/icse/Bertolino07}. Testing can take place at different levels including Unit Testing, which tests individual units of source code; Integration Testing, aimed at testing a collection of software modules seen as a group; System Testing, carried out on a complete integrated system to evaluate its compliance with respect to its functional and/or system requirements; and User Acceptance Testing, meant to determine if the systems meets its requirements in terms of a specification or contract, and allow the user to accept the system, or not. 
\begin{itemize}
\item {\bf Inputs:} Depending on the test level and on the testing technique adopted for the level, the inputs that feed the ST application can change dramatically. For example, Unit Testing tests source code, while System Testing may follow a black-box approach, with the stucture of the system unknown to the tester. 

\item {\bf Outputs:} As for RV, ST always consists of observing a sample of executions, and giving a verdict over them. The output may be an association between a test case, an execution run, and the respective verdict.  

\item {\bf Strengths:} ST can be performed in an agile way \cite{fowler2001agile} and can be directly integrated in the software development method \cite{beck2003test}, allowing the software to be developed and tested in the same development cycle. ST techniques may be very efficient; coverage analysis and reliability measures can help in performing an `as complete as possible' testing.  

\item {\bf Weaknesses:} Software Testing is not exhaustive. Identifying the most relevant units, functions, or functionalities  to test is not easy, and finding the right balance between testing enough, but not too much, requires a deep understanding of the system under test. In particular, testing autonomous systems is hard since 
the range of possible behaviours that can be exhibited by an autonomous agent can be vast~\cite{WinikoffCranefieldJAIR,WinikoffJAAMAS2017}.
The design and development of test cases is expensive; if the system under test changes, test cases developed for a previous version might not be reusable for the new one. Importantly, ST can never verify the absence of bad or undesirable behaviours, only the (partial) existance of desirable ones. Also unlike formal methods, ST requires a complete, executable system to test, pushing ST verification later in the design lifecycle than formal verification, which easily analyzes partial, early system designs. 

\item {\bf Applicability:} While software testing is commonly used in
  industry for both the \lowlayer\ and \midlayer, and is applied
  for testing features at both levels in autonomous
  \cite{helle2016testing,scrapper2006moast} and multiagent systems
  \cite{DBLP:conf/aose/NguyenPBPT09}, we are not aware of proposals
  for testing the \toplayer\ of an autonomous system.
\end{itemize}

The analysis that we carried out shows that no golden bullet exists for addressing the certification, via formal and semi-formal verification, of autonomous systems. All the methods considered in this section show strengths and weaknesses.  
Also, the applicability to the \toplayer\ is not widespread yet. While a few proposals exist for some formal methods, 
the challenges to face are many and the initial ideas sprouting from the research community must still consolidate and turn into practical and usable applications.


\section{Illustrative Examples}
\label{sec:cases}


In this section we discuss some examples demonstrating the generality of the framework.
The authors are aware that currently these systems are not likely to be certified; however, this might change in the future. 

We exemplify systems that can operate independently, and argue that we can extend the framework for the case of operation in a group.  For example, we can consider a single autonomous aeroplane, or a swarm of UAS; drive a car, or do car platooning; interact with a stand-along personal assistant agent (PAA), or use this as an interface to a complete smart-home environment. 

Table~\ref{tab:cases-overview} contrasts over several dimensions some examples of autonomous systems in various domains: the same systems/domains were illustrated in Fig.~\ref{fig:matrix}. 
Note that the second column of Table~\ref{tab:cases-overview} indicates the (current or near future) \emph{scope}  of autonomy (based on Fig.~\ref{fig:matrix}), whereas the next column indicates the (somewhat longer-term future) \emph{level} of autonomy. 

We remind the reader of the levels of autonomy introduced in Section \ref{subsec:terminology}, Definition \ref{Def:Level}: no autonomy; low autonomy; assistance systems;  partial autonomy; conditional autonomy; high autonomy; full autonomy.

Our examples are often quite distinct and, in particular, differ with
respect to the complexity of their decision-making. By this, we mean
the amount of information used to determine the system's behaviour in
any given situation.  Usually, this complexity is known only to the
designer of the autonomous system, it can not be observed from the
outside. In the most basic form, decisions are based directly on
inputs of certain sensors such as ultrasonic distance sensors, lidars,
cameras, etc. A typical decision would be ``if there is an obstacle to
the left, then move to the right''.  We consider such low-level
decisions, where the action is derived directly from the sensory
input, to be of low complexity.  Additionally, however, the system can
have some built-in knowledge concerning its environment that
influences its behaviour; this could be map data, time tables,
situation databases, rule books, etc.  The internal state of an
autonomous agent has been modelled in terms of its beliefs, intentions
and desires.  We consider algorithms which rely on extensive models of
the environment to be of medium or high complexity.  More
sophisticated algorithms are taking the history of decisions into
account, thus ``learning from experience''.  Such algorithms can adapt
to changing environments and tasks, evolving constantly.  Thus, even
the designer of a system can not predict its behaviour with certainty,
if the system has been operating for some time.  We consider such
decision making to be of very high complexity.

The examples differ also with respect to their potential harm, here
defined generally as ``negative outcomes'' to individuals or
society. One aspect of this is safety --- the absence of physical
harm.  In the international standard IEC 61508 on functional safety,
five safety integrity levels (SILs) have been defined.  SIL0 means
there are no safety related requirements, SIL1--4 refer to low,
medium, high and very high safety related requirements, respectively.
%
Other domains such as aerospace and automotive have come up with
similar safety classifications.  The categorization of a (sub-)system
into a certain SIL class is done by considering the risk which the
system imposes on its environment.  This is the likelihood of
occurrence of a malfunction multiplied by its possible consequences.


It is clear that systems/scenarios with higher levels of potential
harm will require strong regulation.  So, both in
Table~\ref{tab:cases-overview} and later subsections, we have
highlighted the level of regulation available. In most cases, this
regulation does not mention `autonomy' nor consider that the human
might not be ``in control''. We would expect that systems with high
potential risk will have stronger forms of regulation.  The
\emph{current} amount of available regulation for each of the sample
systems is indicated in Table~\ref{tab:cases-overview}, ranging from
`low' (scattered or unsystematic regulations; points of inconsistency
or obsolescence in regulations) to `high' (comprehensive regulatory
framework; consistent and current regulations). In addition, once
autonomy is introduced, and especially where this can impact upon
safety or other potential harms, then enhanced regulations for this
aspect will be essential. As yet, specific attention to aspects and
implications of autonomy is rarely made.

\begin{landscape}
\begin{table}[htp]
\caption{Examples of autonomous systems.}
\begin{center}
\begin{tabular}{l ccccc}
\toprule
System  &  Scope of & Targeted Future  & Complexity of    &  Potential   & Amount of\\
       &  Autonomy & Autonomy         &  Decision Making &  Harm        & Existing Regulation\\
\midrule
Robot vacuum cleaner       & low & high  & low & none        & low \\
Autonomous trading system &  low-medium & high & low & high   & low-medium \\
Driverless train                    &  low-medium & full  & medium & very high        & high \\
Unmanned aerial system &  medium & full  & high & very high      & medium \\
Self-driving car                    & medium-high & full  & high & high         & low-medium \\
Personal assistant agent  & high &  full  & very high & low-medium        & low \\
Home assistant robot                 & high & high  & very high & medium        & low \\
\bottomrule
\end{tabular}
\end{center}
\label{tab:cases-overview}
\end{table}
\end{landscape}

In the following, for each example we first describe the functionalities, and
then position the functionalities of the example w.r.t.\@ the three layers in our reference autonomy framework. We then describe the trends in the domain, in particular around the future level of autonomy per Definition \ref{Def:Level}.
Subsequently, we comment on the safety-criticality of the example, and hence the amount of likely needed and available regulation. 
Finally, we comment on suitable verification, validation, and analysis techniques for the example.

\subsection{Simple case: Robot Vacuum Cleaner}\label{sec:roomba}

\noindent\textbf{Functionalities:} 
We begin with a simple example: a domestic robot tasked with simple objectives,
such as vacuuming or sweeping the floor, or cutting grass in the lawn.  

\noindent\textbf{Positioning w.r.t.\@ the layers in our reference autonomy framework:} 
In terms of the three layer model, this sort of domestic robot illustrates that
in some situations, not all of the layers are needed.  There is, of course, a
need for a \lowlayer, which deals with physical aspects of the environment
(sensing, positioning, locomotion, and physical actions).  There is also a need
for a \midlayer\ that follows well-defined rules for `normal' situations. Such
rules might specify a regular schedule for cleaning, as well as indicating when
this schedule might be abandoned or changed (\eg due to human request, or the
system detecting a situation in which cleaning ought to be postponed, such as a
dinner party). The vacuum cleaner must limit its maximum speed to avoid harm,
change direction if it encounters an obstacle, stop if it detects a fault of
any kind, and obey the boundaries of its electric fence.  The \toplayer\ of a
vacuum cleaner is trivial, since the system's functionality is relatively
simple. Specifically, we do not expect the system to deal with any unusual
situations that would require it to override the `normal situation' rules. The
cases when the robot is stuck, or jammed, or too hot due to unexpected warming,
can be considered as `normal situations' and dealt with by the simple `switch
off' rule. The only way the application of this rule could harm, would be to
cause a human to stumble into the robot which stopped in an unexpected
position, but in this scenario we are assuming that humans can cope with the
presence of the robot in their home, which means, they have the ability to
perceive and avoid it. 

\noindent\textbf{Level of (future) autonomy:}
According to our categorization, such robots are designed to operate highly
autonomous in their respective environment.  Human assistance is necessary only
for the initial set up, for emptying the dustbin, refilling cleansing fluids,
maintaining brushes or blades, and repositioning the robot in case it is stuck
or out of battery power.  However, the complexity of decision making in these
devices is low: Often, the control programs are based on a few sensory inputs
only.  Many of these robots navigate the area in a random fashion, turning upon
contact with another object, and moving towards the docking station guided by
an infrared beam if the power is low.  More sophisticated models can learn a map
of their operating environment and use advanced path planning methods to
traverse the whole area.  There are no complex interactions with other devices
or actors in the environment.
The scope of autonomy is also low, given the limited autonomous functionality of these systems.

\noindent\textbf{Safety criticality (of autonomous aspects):}
For this example we exclude the possibility of the tasks having any direct
safety consequence (such as caring for ill humans, having responsibility for
monitoring for injury, or providing medication), and we assume that the humans
interacting with the robot are aware of its existence and can cope with it in a
`normal' way.
Therefore, the safety criticality of the devices is low.

\noindent\textbf{Amount of available regulation (for autonomous aspects):} 
With respect to available regulation, there are just the general laws governing
the use of electric equipment, batteries, etc.

\noindent\textbf{Suitable verification, validation, and analysis techniques:} 
Cleaning robots are often adopted as simple and affordable examples to show the
features and potential of different programming languages, tools, and even
verification and validation techniques \cite{8102947}. Recently, four of the
latest premium robot vacuum cleaners were analyzed and compared by IoT experts
according to security and privacy issues \cite{vacuum-cleaners}, but no formal
techniques were used there. Concerns have arisen since some companies wanted to
use data collected by such home devices --- layout of the house, operating
hours etc. --- for marketing purposes. While we agree that all the techniques
reviewed in Section~\ref{sec33} are suitable for checking that the robot
\lowlayer\ and \midlayer\ behave as expected, their application on real systems
is often considered not to be worth the effort, given the absence of safety
issues. However, these devices make good exercises for academic education in
these techniques.

\subsection{Autonomous Trading System}\label{sec:tradingsystem}

\noindent\textbf{Functionalities:} 
Autonomous trading systems (ATSs) operate in financial markets, submitting buy
and sell orders according to some trading strategy.  ATSs have a history back
to 1949, but came to widespread attention after the `flash crash' on the US
stock exchange in 2010.  By 2014 it was reported that in excess of 75\% of
trades in public US security markets were by ATSs \citep{Levine13:quiet}.  An
ATS submits market instructions, primarily in the form of a buy or sell order.
It may decide when to do so, and the commodity, quantity, and bid or ask price.
ATSs can operate in any market where they are permitted, such as a stock market
or an energy market.  In notable contrast to un-aided human traders, ATSs can
operate high-frequency trading.  The price prediction techniques and trading
strategies vary.  For example, the stock price of a single company might be
predicted using a non-linear function approximator \citep{Bao17:ats}, and the
trading strategy might be based on a set of rules.

\noindent\textbf{Positioning w.r.t.\@ the layers in our reference autonomy framework:} 
With respect to the layers in our reference autonomy framework, 
we can locate the communications protocols and simple but fast
responses to market stimuli (``if the price is below threshold, then issue a
sales request'') at the \lowlayer. More sophisticated trading rules and
strategies, which take general market evolution and developments into account
(``if there is a shortage of oil, then coal prices are likely to rise''), are
located at the \midlayer. 
Although we might anticipate a possible \toplayer\ that reasoned about unusual situations, such as suspending the usual trading strategy in the face of a stock market crash, the speed of trading makes this unlikely, and in practice, we would not expect a \toplayer\ to be used.

\noindent\textbf{Level of (future) autonomy:}
Current ATSs are already capable of operating autonomously, without
moment-by-moment human intervention.  Nonetheless, ATSs can be used also as an
assistant by a human trader, for example to automatically build or
maintain a position specified by the trader.  It is likely that the degree of
autonomous operation and the time-scale of operation will increase, whereas
humans become less able to have oversight of the trading behaviour except at a
high level.
The scope of autonomy depends on the sophistication of the ATS, and varies from low (for relatively simple ATSs that follow simple trading rules), to medium for more sophisticated ATSs.

\noindent\textbf{Safety criticality (of autonomous aspects):}
Given the disruption to financial markets from ATSs, they have a high level of potential harm, and hence are considered safety-critical.  Issues of concern with respect to safety include software errors, herd trading, market fragility, manipulative market strategies, lack of scrutability, the need for an algorithmic `kill switch', and fair market access.

\noindent\textbf{Amount of available regulation (for autonomous aspects):} 
Given the safety criticality of ATSs, the needed regulation is high. 
Regulation has increased since ATSs became a standard tool in the financial
sector in the 21st century.  In the US, the financial authorities introduced
stricter rules following the 2010 crash.  These pertain to algorithmic trading
behaviour, but also to the use of ATSs more generally, and the responsibility
of the companies using ATSs to have adequate supervision of ATSs \citep{FINRA}.

\noindent\textbf{Suitable verification, validation, and analysis
  techniques:} 
  These ATSs are essentially feedback control systems. As such,
standard analysis techniques from control systems such as analytic
stability proofs  
or algorithmic hybrid systems
verification \cite{HenzingerHW97,platzer08:_keymaer}
could be
deployed. However, such approaches require a very good model of the
environment, and its reactions, in which the trading system
resides. 
In particular, the environment includes other ATSs, and due to commercial sensitivity, little is likely to be known about them.
Without a precise environmental formalism, verification and
validation techniques are of little value. Testing can be applied in
this area but, again, strong verification cannot be
achieved. Consequently, there is a danger of `run-away' feedback loops
that are impossible to predict beforehand. 
\later{Need some citations in this paragraph (commented out)}

\subsection{Driverless Trains}\label{sec:trains}

%
%
%
%
%

\noindent\textbf{Functionalities:} 
Automatic railway systems for the transportation of people have been in use since the beginning of the
21$^{st}$ century.  Generally, these systems operate on dedicated routes, which
are protected from human access by special barriers.  Moreover, often
the systems are designed in a way such that a collision of trains is
physically impossible.  Therefore, reliability of these systems
usually is very high, with failures being caused by mechanical fatigue
rather than by software errors. 

\noindent\textbf{Positioning w.r.t.\@ the layers in our reference autonomy framework:} 
All modern rail vehicles have a Train Control and Management System (TCMS), which in conjunction with the on-board signalling computer (OBC) is responsible for the basic operations of the train. With respect to our three-level framework, TCMS and OBC realize the \lowlayer.
In the autonomous subway example described above, the \midlayer\ is responsible for planning the journey, stops in the station, etc.
A \toplayer\ would be needed at least for unattended train operation, and even more for trains which calculate their own route on the tracks. 
There are many `unusual' situations that can arise in such a scenario, which have to be dealt with by an \toplayer.

\noindent\textbf{Level of (future) autonomy:}
Most present-day driverless trains, e.g., people-mover in airports, have almost no autonomy.
They operate according to fixed time schedules, with fixed distances to other trains, and under tight supervision of a human operator.
The situation becomes more interesting when looking at self-driving underground vehicles.
Some of these trains can also drive on tracks used by conventional subway trains.
Thus, the autonomous vehicle has to decide when to advance, and at which speed. 
In N\"urnberg, such a system has been operating for over a decade, with more than 99\% punctuality and without any accidents.
A main benefit of the autonomous operation is that it allows for a higher train frequency than with human-driven trains.
In underground train systems, the probability of an obstacle on the
tracks is rather low.  For trams, regional trains, or high-speed
trains, this is different.  People or animals may cross the tracks, and thunderstorms
may blow trees onto the tracks.  Therefore, one of the main tasks of a
train driver is to supervise the journey and to brake if necessary.
Other tasks include monitoring for unusual events, checking the
equipment before the trip, and passenger handling.  There is an
increasing number of electronic driver advisory systems such as speed
recommendation systems, driver information systems, and others.
Currently, all these systems leave responsibility for the trip to the
driver.  However, there is no compelling argument that an automated
system could not do this task, increasing the autonomy of the system even further.
However, the scope of autonomy is not high, since the functionality of the system is constrained by its environment (running on tracks).

The norm IEC 62290-1:2014 (Railway applications -- Urban guided transport management and command/control
systems -- Part 1: System principles and fundamental concepts) defines
five grades of automation (GoA).  
These can be compared with the levels of autonomy given in Definition \ref{Def:Level}.
The fourth grade is driverless train
operation (DTO), where the only tasks of the (human) train attendant are to handle passengers,
control the doors and to take over control in emergency situations.
The fifth grade is unattended train operation (UTO), where there is no
personnel from the train line operator on board.  That means, the
electronic systems need to be able to deal also with emergency
situations.

%

\noindent\textbf{Safety criticality (of autonomous aspects):}
Compared to other domains, railway homologation and operation is strictly regulated.
For the approval of a particular system, the authorities rely on notified bodies, designated bodies and associated bodies.
These are institutions that assess the conformity of rail equipment to certain standards such as Cenelec~50128~\citep{cenelec50128}.
The assessment processes are strongly schematic.
For safety-critical software, known best practices, including formal methods, have to be applied.

\noindent\textbf{Amount of available regulation (for autonomous aspects):} 
However, the available regulation in the railway sector is not prepared to deal with higher levels of autonomy.
For example, the movement authority, \ie the permission to enter a certain block, is currently
given by the roadside equipment or the railway control centre.  There
are research projects investigating whether it is possible to transfer
this responsibility to the train; this would require safe localization
of a train and reliable communication between trains.  In
such a truly autonomous setting, the `electronic driver' not only
has to ensure that the track is free from obstacles, but also that
there are no other trains in its way.
Currently, it is unclear how such a system could be officially approved.

\noindent\textbf{Suitable verification, validation, and analysis techniques:} 
Railway software must be extensively tested and verified before it can be put into use.
Due to the cost and complexity, formal verification is done for functions of the highest criticality only.
These are mainly the signalling and breaking functions.
Testing is done on unit, integration, and system level: 
in contrast to the automotive domain, road tests are extremely expensive; thus, developers strive to limit the necessary road tests to a minimum.
For autonomous driving, the problem becomes dramatically more severe, since it is not possible to test millions of different situations on the track. 


\subsection{Unmanned Aerial System}\label{sec:uas}


\noindent\textbf{Functionalities:} 
The term `Unmanned Aerial Systems' (UAS) covers a wide range of airborne vehicles, from toy quadrocopters to military drones and autonomous missiles.
Whereas for transportation of people an airplane without pilot is (yet) beyond discussion, the transport of goods and material by UASs is a thrilling perspective.
However, the main public concern is that a falling UAS might harm people on the ground.

\noindent\textbf{Positioning w.r.t.\@ the layers in our reference autonomy framework:} 
The mapping to our three layer framework is fairly clear, as is the need for all three layers. 
Rapid responses to the physical environment are provided by the \lowlayer, and there is a need to be able to apply rules to usual situations (\midlayer) as well as deal with unanticipated situations by applying principles (\toplayer).

\noindent\textbf{Level of (future) autonomy:}
Already now, human pilots have a wealth of support systems at hand: navigation systems, autopilots, traffic collision avoidance systems, etc.
Vital flight information is recorded and transmitted to ground control.
Cameras are available providing a visual image from the cockpit.
Therefore, moving from a human pilot within an air vehicle, to a human pilot
remotely controlling the vehicle, and then towards the vehicle
controlling itself autonomously, is a very popular
approach. Yet it is fraught with difficulties (as discussed in earlier sections, in particular Section~\ref{sec:where-we-are-now}).
In the near future we expect the scope of autonomy to be somewhat limited (medium), whereas it is possible that longer-term we will see UASs with higher levels of autonomous functionality.

\noindent\textbf{Safety criticality (of autonomous aspects):}
Even though we are not (yet) considering UAS that carry human passengers, the safety criticality of UAS is very high. This is for the simple reason that a significant malfunction might lead to a collision, either between the UAS and a ground object (e.g.~building, people), or between a UAS and another aircraft (potentially carrying passengers).

\noindent\textbf{Amount of available regulation (for autonomous aspects):} 
To provide a practical
route to certification, including for non-autonomous systems, a range of
documents/standards have been provided~\cite{PatchettJumpFisher15},
perhaps the most relevant being the FAA documents \emph{Software Considerations in
  Airborne Systems and Equipment
  Certification}~\cite{DO-178B,DO-178C}, \emph{Formal Methods Supplement to DO-178C and DO-278A}~\cite{DO-333}, and \emph{Design Assurance Guidance for Airborne
Electronic Hardware}~\cite{DO-254}. These standards provide
directions toward the use of formal methods in the certification of
(traditional) air systems, but say relatively little about autonomous
(or even semi-autonomous) operation. The FAA has recently published official rules for non-hobbyist small UAS operations, \emph{Part 107 of the Federal Aviation Regulations}~\cite{Part107} that certifies remote human operators for a broad spectrum of commercial uses of UAS weighing less than 25 kg.

\noindent\textbf{Suitable verification, validation, and analysis techniques:} 
One approach, discussed in Section~\ref{sec:process},  to the analysis and potential certification of unmanned,
particularly autonomous, air systems is to begin to show that the
software replacing the pilot's capabilities effectively achieves the
levels of competence required of a human pilot. For example, if we
replace the `flying' capabilities of a pilot by an \emph{autopilot}
software component, then we must prove the abilities of the autopilot
in a wide range of realistic flight scenarios. While this corresponds
to the operational/control \lowlayer\ within our
approach, we can also assess the decision-making against the required
rules that UAS should follow. Pilots are required to
pass many tests before being declared competent, with one such being
knowledge of the `Rules of the Air'.  In~\citep{DBLP:journals/jacic/WebsterCFJ14} the agent
making the high-level decisions in an UAS is formally
verified against (a subset of) the `Rules of the Air'.
This type of analysis of the system's \midlayer\  is
necessary, though not sufficient, to provide broader confidence in
autonomous air systems. Human pilots possess strong elements of
`airmanship' and the ability and experience to cope with unexpected
and anomalous situations. It is here that the \toplayer\ must capture this, more complex, decision-making. An initial attempt
at the formalisation and verification of ethical decisions in
unexpected situations was undertaken in~\cite{DennisFSW16} wherein a
simple ethical ordering of outcomes was used to cope with situations
the UAS had no expectations of, and rules about.

\subsection{Self-Driving Car}\label{cars}

\noindent\textbf{Functionalities:}  
Currently, the development of self-driving cars is one of the main innovation factors (besides e-mobility) in the automotive industry.
Advanced driver assistance functions such as adaptive cruise control, lane keeping, or parking, are available in many cars.
Combining such driver assistance systems, a partial autonomous driving (e.g., driving on the highway for a limited time) can be realized.
Although not yet being available to end customers, cars with conditional autonomy (e.g., driving on the highway for an extended distance or autonomous parking in a parking deck) are already state of the art.
Every major manufacturer has concrete plans to  bring highly automated cars to the market in the next few years, although there is some debate as to whether these plans are feasible.
Intermediate steps towards increasing autonomy include
vehicle platooning/convoying whereby vehicles commit to being part of
a (linear) formation and cede control of speed/direction to the
platoon/convoy leader~\citep{Swaroop1997,bergenhem2010,EU:satreProject}.

\noindent\textbf{Positioning w.r.t.\@ the layers in our reference autonomy framework:}  
In terms of our three-level categorisation, current `driverless'
cars essentially only have the lowest operational/control \lowlayer\ to
handle \emph{lane following}, \emph{speed control}, etc, while the
human driver is responsible for following the rules and regulations
concerning road use and for coping with unexpected situations or
failures. 

Once we move to the platoon/convoy (or `road train') concept, the
aim is that more of the (developing)
regulations~\citep{PlatoonLegislation} are taken by the autonomous
system. Again, the human driver will be responsible for unexpected
situations and failures, and will also take over responsibility for
road use once the vehicle leaves the platoon/convoy.
Thus, in this case, a \midlayer\ is required.

However, the fully autonomous vehicles that we expect in the future will
require software to be able to handle not just operational aspects
such as steering, braking, parking, etc, but also comprehensive
regulations such as the ``rules of the road''. Furthermore, the
\toplayer\ of our autonomous system must conform to some high-level
requirements concerning unexpected situations. Of course these do not
exist yet, but we might expect them to at least have some quite
straightforward basic principles such as
\begin{quote}
  \emph{avoid harm to humans where possible}
\end{quote}
and
\begin{quote}
  \emph{if a problem occurs, safely move to an approved parking area}
\end{quote}
Of course, many more principles/properties will surely be needed.

\noindent\textbf{Level of (future) autonomy:}
A long-term vision is that self-driving cars are available on demand.
If some people need a transport service, they can call an autonomous taxi via a smartphone app.
Thus, there is no necessity to buy and maintain one's own car; resources are shared between many users.
For this vision to become a reality, cars must be able to drive fully autonomous to any desired destination in all traffic situations, and must moreover be able to collaborate with the rest of the fleet to deliver an optimal service.
As this vision is gradually realised, self-driving cars will have increasing scope of autonomy (eventually high), but in the nearer term we expect more limited scope of autonomy.

\noindent\textbf{Safety criticality (of autonomous aspects):}
Clearly, cars have a high level of safety criticality. 
The development of automotive software must follow
functional safety standards, such as ISO 26262.
This standard also prescribes the safety requirements to be fulfilled in the design,
development and validation. 
For the verification and validation of software functionalities on the \lowlayer\ and \midlayer, we regard the existing techniques and regulations as being sufficient.

\noindent\textbf{Amount of available regulation (for autonomous aspects):} 
While there is considerable hype suggesting that such vehicles are imminent, neither the automotive industry nor the regulators are at the point of being able to certify their reliability or safety.
Global legal constraints, such as the Vienna Convention on Road Traffic~\citep{vienna68},
ensure that there must always be a person which is able to control the safe operation of the vehicle.
It is a current political discussion how to amend these regulations. 
The US Department of Transportation has recently issued a ``Federal Automated Vehicles Policy''. It includes a 15-Point Safety Assessment, including ``Testing, validation, and verification of an HAV system''.
For SAE levels 4 and 5,  if the driver has handed the control to the vehicle, liability rests with the manufacturer.
Thus, most car manufacturers follow a ``trip recorder'' approach: A black box specially protected against tampering is used to collect all relevant data, and can be used to assign liability in case of an accident.
However, with current legislation liability rests on factors such as human reaction times, weather conditions, etc.
It is an open question whether these factors can be transferred to the case that an autonomous agent is the driver instead of a human.

\noindent\textbf{Suitable verification, validation, and analysis techniques:}
 Concerning the extraction of requirements and the verification of
 these requirements on a specific vehicle, we can see how the
 operational/practical constraints will be derived from many of the
 existing requirements on \emph{cruise control}, \emph{lane
   following}, etc, while we will see refined requirements for the
 sensing/monitoring capabilities within the vehicle. Verifying all
 these aspects often involve a suite of (acceptable) testing
 schemes. In addition, it is clear that, just as the agent controlling
 an autonomous air vehicle can be verified with respect to the ``rules
 of the air'' (see above), the agent controlling an autonomous road
 vehicle can be verified against the ``rules of the
 road''~\citep{Alves2020}.

However, when it comes to the functionalities of the \toplayer,
the rules that a human driver is expected to follow 
need to be transformed into requirements
suitable for an autonomous agent. This will also include
sensor/camera reliability and confidence in obstacle/sign/pedestrian
recognition, and will also feed in to risk analysis component
within the agent. Example for such `\toplayer\ rules' are
  \emph{``if there is low perceived risk of collision, continue at the highest admissible speed''},
and
\emph{``if there is high perceived risk of collision, or low confidence in the perceptions, then slow down''}.
%
%
In fully autonomous vehicles, the reliability of the \toplayer\ is a crucial aspect.
It must be able to cope with unexpected events, failures, aggressive road users, etc.
Thus, the extraction of suitable requirements and their validation is highly important.

\subsection{Personal Assistant Agent} 
\label{sec:paa}

%

\noindent\textbf{Functionalities:} The now-ubiquitous personal
assistant agent has origins earlier than the web
\citep{Hodgkins11:siri,DBLP:journals/internet/HuhnsS98d} and became
mainstream with Apple's \sysname{Siri} in 2010
\cite{DBLP:journals/tist/BerryGPY11}.  The vision for such a
\defn{PAA} agent is proactive, personalised, contextual assistance for
a human user \citep{DBLP:conf/atal/Yorke-SmithSMM09}.  A PAA agent is
a software entity, although the services it connects to and can
command may include physical services such as smart home devices.  So
far, PAAs are mainly voice-controlled assistants that react to user
requests such as looking for information on the web, checking the
weather, setting alarms, managing lights and thermostats in a smart
home. Few apps classified as smart personal assistants connect to
service providers, banks, social networks; 24me is one among them:
according to its description on the web \cite{24me}, ``automatically
tells you what you need to do and also helps you take care of your
things from within the app''.  This technology moves on rapidly,
however, and Gatebox's ``Azuma Hikari''
system\footnote{\url{https://www.gatebox.ai/en/hikari}} promises a
startling development:
\begin{quote}
  \emph{``The Comforting Bride: Azuma Hikari is a bride character who develops over time, and helps you relax after a hard day.
    Through day-to-day conversations with Hikari, and her day-to-day behavior, you will be able to enjoy a lifestyle that is more relaxed''.} [\url{https://www.gatebox.ai/en/hikari}]
  \end{quote}


\noindent\textbf{Positioning w.r.t. the layers in our reference
  autonomy framework:} The \lowlayer\ for a PAA consists of the PAA
connecting to local and web services.  The services can affect the
world and the user's day-to-day life in both smaller and larger ways:
from opening an app on a mobile device, to reserving a table at a
restaurant; from changing the temperature in the user's house to
making a bank payment from her account. 
The \midlayer, currently only realized
in sophisticated systems, is responsible for the combination of such services,
e.g., according to user preferences (``one hour after the scheduled dinner
reservation, begin heating the house'').
The \toplayer\ would be not only
responsible for exceptional and emergency circumstances (e.g., calling help
when the human needs it), but also for giving advice in personal matters (e.g.,
whether to sign an insurance contract or not).
The ``Azuma Hikari'' character
has the potential for quite an array of autonomous behaviour but, not
just setting out required behaviours, in \midlayer, but even providing
quite sophisticated \toplayer\ behaviours.


\noindent\textbf{Level of (future) autonomy:} The level and scope of autonomy are both
limited so far, but we see a huge potential for PAAs to become more
and more autonomous, implementing Negroponte's vision of the agent
which \emph{``answers the phone, recognizes the callers, disturbs you
  when appropriate, and may even tell a white lie on your
  behalf. [...] If you have somebody who knows you well and shares
  much of your information, that person can act on your behalf very
  effectively. If your secretary falls ill, it would make no
  difference if the temping agency could send you Albert
  Einstein. This issue is not about IQ. It is shared knowledge and the
  practice of using it in your best interests''}
\citep{Negroponte:1996:DIG:223799}. Again, the ``Azuma Hikari''
approach provides the possibility for quite a lot more autonomous
behaviour.

\noindent\textbf{Safety criticality (of autonomous aspects):} 
Depending on what the PAA can control, it may actually be a safety critical system: if it can make financial transactions, it might unintentionally cause economic loss to its user; if it can send messages on behalf of, or even pretend to be its user, it might create severe damages to professional and personal relations. Indeed, if the PAA is given control over aspects of the user's physical environment (such as locks on their home's doors), the possibilities for harm become more significant.
%
While not safety-critical, the ``Azuma Hikari'' assistant clearly has
some potentially ethical issues~\citep{Pietronudo2018JapaneseWL} as
captured in the BS8611, not least with the anthropomorphism. If we
extend our view of safety beyond physical harms to more general
psychological/ethical harms then there might well be issues here.

\noindent\textbf{Amount of available regulation (for autonomous aspects):} 
 To our knowledge, there is no regulation of PAAs, other than general consumer electronics regulation and regulations in specific sectors which co-incidentally apply to the PAA's services such as regulations concerning protocols for accessing banking institution services. While there is no single data protection legislation in the US, in Europe the General Data Protection Regulation (GDPR \citep{GDPR2018}) became operative in  May 2018. GDPR regulates the processing by an individual, a company or an organisation of personal data relating to individuals in the EU. A PAA based in Europe should comply to the GDPR. 
 As an example, it should not share information by making unauthorized unauthorized calls on behalf of its user and should not share sensitive data (pictures, logs of conversations, amount of money in the bank account, health records) of its user and his or her relatives. It should also take care of where these personal data are stored to ensure an acceptable level of protection. On the other hand, a PAA should be allowed to override privacy laws and call emergency services if it suspects its user is unconscious from an accident, suicidal, or otherwise unable to ask for help, and it should share health information with the first aid personnel. Relating to the ethical/psychological issues above,  standards such as BS8611 and the IEEE's P7000 series are of relevance.

 {\bf Suitable verification, validation, and analysis techniques:}
  Like any complex piece of software, the PAA should undergo a careful testing stage before release. However, we believe that the most suitable technique to prevent it from unsafe behaviour, is runtime verification. The financial transactions might be monitored at runtime, and blocked as soon as their amount, or the amount of the exchanged message, overcome a given threshold. Calls to unknown numbers should not necessarily be blocked, as they might be needed in some emergency cases, but they should be intercepted and immediately reported to the user. Commands to  smart devices should be prevented, if they belong to a sequence of commands meeting some dangerous pattern.

\subsection{Home Assistant Robot}\label{sec:homehelp}

The situation changes if we consider home assistant robots that are not only
performing simple tasks, but also issuing medical reminders, providing simple
physiotherapeutic treatment, and even offering `companionship'.  Consider a
robot such as the `Care-o-Bot'\footnote{\url{http://www.care-o-bot.de}}.  A number of
domestic assistants are already available. These span a range, from the above
Robot Vacuum Cleaner right up to Care-o-Bot systems, and similar systems, such
as Toyota's Human-Support Robot Family\footnote{\url{www.toyota-global.com/innovation/partner_robot}}. 
We will primarily
consider the more sophisticated end of this spectrum.


{\bf Functionalities:} 
Robots such as Care-o-Bot are
intended to be a general domestic assistant. They can carry out menial
cleaning/tidying tasks, but can also (potentially, given the regulatory
approval) fetch deliveries from outside the home and provide food and drink for
the occupant(s) of the house. Each such robot is typically mobile (\eg wheels)
and has at least one manipulator, usually an `arm'. Yet this type of robot can
potentially go further, becoming a true \emph{social robot}, interacting with
the occupant(s) (\eg what to get for lunch?) and engaging in social dialogue
(\eg favourite TV show?). These robots (again, with appropriate approval) also
have the potential to measure and assess physical and psychological health.
Clearly, with this wide range of functionalities and increasing autonomy,
robots of this form should be both highly-regulated and will have many,
non-trivial, decisions to make.

{\bf Positioning w.r.t.\@ the layers in our reference autonomy framework:}
The need for all of our layers is clear. Any physical interaction
with the environment, or movement within a house, will require a complex
\lowlayer. Similarly, one would expect the regulatory framework for such
robotic systems to be quite complex and so, necessarily, the  \midlayer\ will
also likely be comprehensive. In the future the combination of broad
functionality, increasing autonomy, and responsibility for health and
well-being of human occupants will surely lead to quite complex ethical
issues~\cite{Anderson2:08,Matthias:childrenRI,SalemLAD15:2,2017arXiv:170304741C}
and so to a truly non-trivial \toplayer.

{\bf Level of (future) autonomy:}
In principle, the scope and level of autonomy
can both be high but, as noted below, the regulatory frameworks limit this at
present.  In the future, home care robots might become much more than service
providers. Especially with people who are isolated and find little human
contact, these robots might begin to guide and accompany their owners, remind
them to take their medicine or drink water, and even provide simple social
companionship (for example, asking about relatives, about the news, or about TV
programmes seen recently).  Especially in these latter cases, the issue of
\emph{trust} is crucial~\cite{amirabdollahian2013can,SalemLAD15,Moral:FLoC18},
since no one will use a robot in this way if they do not trust them. It becomes
a central issue for the {\toplayer} to assess how much to build trust and how
much to stay within legal/regulatory bounds. Doing what the owner wants it to
do can build trust, whereas refusing to do something because of minor
illegality erodes trust very quickly. 
\later{need citation here ref:trust2}

{\bf Safety criticality (of autonomous aspects):}
In addition to the various issues that apply to Personal Assistant Agents, the broader role of Home Assistant Robots makes them more safety critical. For instance, if medicine is administered incorrectly, or if a Home Assistant Robot failures to raise the alarm when a human has an adverse health event, then lives may be put at risk.
Therefore, it appears that clear
regulations specifically developed for such robots will be needed, alongside
standards targeting exactly the issues concerning these installations. With
the complex practical, social, and health-care potential, the range of
standards needs to go beyond physical safety and on to issues such as ``ethical
risk''~\cite{BS8611}.

{\bf Amount of available regulation (for autonomous aspects):} 
There appears to be very little
current regulation specifically designed for autonomous robotics, let alone
such autonomous domestic assistants. While there are some developing standards
that have relevance, such as ISO~13482~\cite{ISO13482}, the regulatory
oversight for domestic robotic assistants is unclear. In some cases they are
treated as industrial robots, in some cases as medical devices (especially if
health monitoring is involved), and in others they are straightforwardly
treated as electro-mechanical devices. In some jurisdictions (\eg the UK)
responsibility for regulation falls to the local regulatory authority, whereas
in others regulatory oversight is a national responsibility. Furthermore, the
standards that are appealed to are not specific standards for autonomous
domestic robots, but rather standards for software, robots, or medical
appliances.
    
On particular challenge concerns learning new behaviours. These might be specified by the
user\later{need citation here}~or might result from analytical analysis by the robot. These
learned behaviours might well be `new', \ie not part of the robot's behaviour
when initially deployed. In this case, there should be some mechanism to assess
whether the new, learned behaviour is consistent with the prescribed
regulations. However, this assessment must either take place offline, in which
case there will be some `down time' for the robot, or online, in which case we
will need effective and efficient techniques for carrying out this
assessment. \later{need citation here} In the case where the human has specified/requested
some new behaviour that is `in conflict' with the regulations, the {\toplayer}
becomes essential as it must make decisions about whether to employ, or ignore,
this new behaviour.

{\bf Suitable verification, validation, and analysis techniques:}
Ignoring the analysis of standard robot manipulation and movement aspects,
there has been specific work on the analysis of
\lowlayer~\cite{WoodmanWHF12,KhanHPMS10} and
\midlayer~\cite{TAROS14:fridge,THMS15} issues. As we would expect, the ethical
issues concerning such (potential) robots are of great interest to both
philosophers and formal
methods~\cite{Anderson2:11,Moral:FLoC18}. Interestingly,
issues such as privacy and security are only gradually being considered
here~\cite{XiaoLD08}


\section{Future Challenges}
\label{sec:challenges}

This section looks ahead. Based on the discussion in previous sections, it defines a number of challenges. These are broken down into three areas, depending on the intended audience, \ie who are the people who could make progress on the challenge? Specifically, we consider challenges to researchers, to engineers, and to regulators.
%

\subsection{Research challenges}

There are a number of challenges that we identify for the research community. 

One group of challenges relates to the overarching question of how to extend from current verification settings and assumptions, to the settings and assumptions required? Specific questions include:
\begin{itemize}

\item As described earlier (Sections~\ref{sec:cando} and~\ref{sec:cases}), in various domains autonomous systems need to perform ethical reasoning. However, how to specify the reasoning required, and perform it effectively, is still an open question, as is how to specify verification properties that relate to such reasoning.

\item In domains that involve interaction with humans, especially close interaction, such as human-agent teams, there is a need for ways to elicit and model the capabilities and attitudes (\eg goals, beliefs) of the humans.

\item Since autonomous systems, even in safety-critical situations, are increasingly likely to make some use of Machine Learning, how can we verify such systems? If a system is able to learn at run-time, then some sort of run-time verification will be needed, although there may also be a role for mechanisms that limit the scope of the learning in order to ensure that important desired properties cannot be violated.

\item How can we handle larger and more complex systems (scalability of verification methods)? This includes systems with potentially not just multiple but \emph{many} agents.

\item Finally, how we can improve the way that we design and build systems to make them more amenable to verification? One possible approach is to use \emph{synthesis} to build systems that are provably correct from verification artefacts.
\end{itemize}

In addition to challenges relating to extending verification to richer, and more complex, settings, there are also a number of research challenges that relate to the broader social context of verification.
Perhaps most fundamentally, we need better methods for systematically identifying (\ie deriving) properties to be verified. We have sketched (in Section~\ref{sec:process}) an outline of such a process, but more work is required to flesh this out, and refine it through iterated use and evaluation. 
One particular challenge is dealing with unusual situations that are particularly hard to anticipate. 
Another area that poses research challenges is  the possibility of using natural language processing to create structured formal (or semi-formal) requirements from textual artefacts (which we discussed earlier, in Section~\ref{subsec:standards2spec}).

There is also a big research challenge to develop ways to engineer systems that are transparent, and, in particular, systems that can explain themselves to users, regulators etc.  While there is the whole subfield of explainable AI (XAI), here we highlight the particular needs of safety-critical autonomous systems -- where explainability might need to focus less on explaining specific individual behaviours to users, and more on explaining how a system operates to regulators. 

%
%
%
%
%
%
%
%
%
%
%
%

\subsection{Engineering challenges}

We now turn to engineering challenges. These are challenges that need to be addressed to practically engineer reliable autonomous systems, once the underlying research challenges have been adequately addressed.

\begin{itemize}

\item An existing challenge that is exacerbated is the need to track assumptions and their provenance, and to manage the system as it evolves (maintenance). If requirements change, and the system is modified, how can we update the verification (as well as manuals and other documentation)?

\item We identified research challenges relating to extending the scope and setting to deal with ethical reasoning, and to deal with machine learning.  Systems that perform ethical reasoning must be built; building systems that involve learning also has engineering challenges.

\item A research challenge that we identified was how to model humans, for example in the context of human-agent teamwork. There is a related engineering challenge which is how to design systems that `work well' with humans.  Note this is a design challenge, not a research challenge (there is a body of \eg HCI research on this).

\item Another research challenge was using \emph{synthesis} to create provably correct systems. A related engineering challenge is managing this process. In practical terms, even if we have technology that can generate a system (or parts of a system) from verification artefacts, there are engineering challenges in managing this process, and having a meta-system where the user clicks `go' and the system is built from V\&V artefacts.

\item We noted above a number of research challenges that relate to the broader social context. There are also engineering challenges that relate to this. The big one is linking verification of design artefacts to the broader needs of stakeholders, in particular regulators. Suppose that a software system for controlling a train is verified. How can this be used to provide regulators with a convincing safety argument? One particular aspect is trust in the process and tools: how do we know that the verification tools are themselves correct? (see the discussion of regulator challenges below)

\item Finally, just as there are research challenges in scaling to handle larger and more complex systems, so too there are engineering challenges that relate to verifying large and complex systems. These concern techniques for making effective use of computing resources (be they traditional high-performance computing, or cloud computing) in order to effectively verify systems, and manage this process.

\end{itemize}


%


%
%


Along with these challenges, we mention the difficulty that companies can have in finding qualified engineering staff \citep{forbes17:lack-engineers}.

\subsection{Regulatory challenges}\label{sec:regulatorychallenges}

Finally, there are challenges for regulators. 
A number of challenges relate to the existence of multiple stakeholders, who may be in a complex state that combines cooperation (\eg to create standards) and competition. Related challenges to this include: how to manage disclosure of information that is sensitive (\eg industrial `secrets'), and how to reach consensus of involved stakeholders, \eg car OEMs. 
This is particularly important if autonomous systems from different manufacturers have to collaborate, e.g., in car platooning.

More broadly, the regulatory landscape is complex, with multiple actors (governments, courts, companies, etc.) which poses challenges around how to manage differences between jurisdictions (or between regulators with overlapping domains of interest). 
If autonomous systems operate in different countries, they should comply to all national regulations of the involved countries.
However, the legal requirements may differ, and even contradict each other.
Here, multi-national agreements and treaties are necessary.

When we consider any form of ethical reasoning done by autonomous systems, then there is a challenge in how to obtain sufficient agreement amongst various stakeholders (\eg government, civil society, manufacturers) about the specification of what is appropriate ethical behaviour, or even a clear delineation distinguishing between behaviour that is clearly  considered ethical, behaviour that is clearly considered unethical, and behaviour where there is not a clear consensus, or where it depends on the underlying ethical principles and framework.

A related point is the legal notion of responsibility, which is a question for regulators, and for society more broadly in the form of governments and legal systems; see for instance \cite{Gunkel14:morality-issue} for an introduction.

Finally, as mentioned above, there are challenges in trusting tools and processes themselves. How can the verification tools be trusted? \citep{DBLP:conf/atva/Shankar08}  To what extent can they be verified? \citep{Ramesh01:requirementstraceability}.



\section{Conclusion}
\label{sec:conc}

Autonomous systems have great potential to transform our world. The substantial adoption of even just one type of autonomous system --- self-driving cars --- will substantially alter passenger transport and the geography of cities (e.g.,~enhancing mobility for those who cannot drive, and reducing the need for parking spaces). More fundamentally, autonomous systems change our relationship with technology, as technology and human society reach an ``intimate entanglement'' \citep{DBLP:journals/cacm/FrauenbergerP19}. 
However, ensuring that autonomous systems are fit-for-purpose, especially when their function is in any way safety-critical, is crucial for their adoption. 
This article therefore addressed the fundamental question: ``\textit{How can the reliability of an autonomous software system be ensured?}''  

After reviewing the state-of-the-art, including standards, existing approaches for certification, and open issues, this article proposed a way forward towards a framework for  certification of safety-critical autonomous systems. We presented  a three-layer framework for structuring such systems, gave an indication of what is needed from regulators, and outlined a process for identifying requirements. We  reviewed a range of verification techniques, considering their applicability, strengths and weaknesses with respect to autonomous systems, and illustrated the application of our framework in seven diverse application scenarios. 
Finally, in order to help move towards a detailed and usable framework for certification, we articulated a range of challenges for researchers, for engineers, and for regulators. 

In addition to the specific challenges discussed in the previous section, there are also a number of other questions arising from the emergence of autonomous systems. These are cross-cutting, in that they pose challenges to all of the key stakeholders (researchers, engineers, and regulators).
\begin{itemize}
\item How to deal with tacit knowledge, e.g.,~where humans learn by ``feel'' (e.g.~pilots)? Should each individual autonomous system learn, or should the knowledge evolve in a population?
\item How to deal with security challenges, in particular, if the autonomous system cannot be continually supervised? Should an autonomous system have ``self-defence'' capabilities against humans?
\item How to deal with Quality of Service (QoS) requirements, in particular, for a large group of autonomous systems and many users?
\item How to deal with varying interpretations, inconsistent requirements, and missing context?
\item How to deal with contradicting requirements from different stakeholders? Is there a notion of ``loyalty'' for autonomous systems? 
\item How to deal with changing standards, attitudes, morals?
\end{itemize}


We hope that the initial framework, and the specific challenges, can form a map in guiding the various communities along the path towards a framework for certification of reliable autonomous systems.


\begin{paragraph}{Acknowledgements.}
We thank the organisers and participants of the Dagstuhl~19112
workshop.  Thanks to Simone Ancona for the drawings in Section
\ref{sec:intro}. Work of Fisher supported by the Royal Academy of
Engineering and, in part, through UKRI ``Robots for a Safer World''
Hubs EP/R026092, EP/R026084, and EP/R026173. Work of Rozier supported
in part by NSF CAREER Award CNS-1552934, NASA ECF grant NNX16AR57G,
and NSF PFI:BIC grant CNS-1257011.  
\end{paragraph}



\later{Clean up bibliography; including ``NSDI'' for Musuvathi et al, and avoiding ``and others''; also URL layout and avoiding both DOI and URL?}



\end{document}